\documentclass[aps,pra,preprint,preprintnumbers,amsfont,amsmath,amssymb,showpacs,a4paper,twoside]{revtex4}

\usepackage[dvips]{graphicx}
\usepackage[flushmargin]{footmisc}
\usepackage{psfrag,epsfig,bm,dcolumn,color}
\def\PJ#1{}
\renewcommand{\PJ}[1]{{\color{red}{\tt ***PJ*** #1}}}
\def\thecomma{\ifx,\thenext \else\ifx;\thenext \else\ifx. \thenext \else\ifx!\thenext \else\ifx:\thenext \else \  \fi\fi\fi\fi\fi}
\def\condblank{\futurelet\thenext\thecomma}
\def\ie{{i.e.,} }
\def\eg{{e.g.,} }
\def\viz{{viz.,} }
\def\rhs{{r.h.s.}\condblank}
\def\lhs{{l.h.s.}\condblank}
\def\fig{{Fig.}\condblank}
\def\slim{-4.5mm} 
\def\tlim{-1mm} 
\def\scale{0.65} 
\def\p{\;} 

\begin{document}

\title{Time delay for one-dimensional quantum systems\\ with steplike potentials}

\author{W.~O.~Amrein}
\email[E-mail: ]{werner.amrein@physics.unige.ch}

\author{Ph.~Jacquet}
\email[E-mail: ]{philippe.jacquet@physics.unige.ch}

\affiliation{D\'{e}partement de Physique Th\'{e}orique, Universit\'{e} de Gen\`{e}ve, CH-1211 Gen\`{e}ve 4, Switzerland.}

\received{\today}

\begin{abstract}

This paper concerns time-dependent scattering theory and in particular the concept of time delay for a class of 
one-dimensional anisotropic quantum systems. These systems are described by a Schr\"{o}dinger Hamiltonian $H = -\Delta + V$ 
with a potential $V(x)$
converging to different limits $V_{\ell}$ and $V_{r}$ as $x \rightarrow -\infty$ and $x \rightarrow +\infty$ respectively. Due to the anisotropy they exhibit a two-channel structure. We first establish the existence and properties of the channel wave and scattering operators by using the modern Mourre approach. We then use scattering theory to show the identity of two apparently different
representations of time delay. The first one is defined in
terms of sojourn times while the second one is given by the
Eisenbud-Wigner operator. The identity of these representations is well known for 
systems where  $V(x)$ vanishes as $|x|
\rightarrow \infty$ ($V_\ell = V_r$). We show that it remains true in the anisotropic case $V_\ell \not = V_r$, \ie  we prove the existence of the
time-dependent representation of time delay and its equality with
the time-independent Eisenbud-Wigner representation. Finally we use this identity to give a time-dependent interpretation of 
the Eisenbud-Wigner expression which is commonly used for time delay in the literature.

\end{abstract}


\maketitle

\section{Introduction}\label{Introduction}

Time delay is an important concept in scattering theory. In the simplest situations it expresses the excess time that scattered particles spend in the scattering region when compared to free particles subject to the same initial conditions. A positive time delay means that particles take more time to pass through the region where they are influenced by the interaction than particles propagating freely through the same region. A negative time delay means that on average the scattered particles are accelerated by the effects of the interaction.

Different approaches to the definition of time delay and related concepts, together with various applications to physical problems and a considerable number of references, have been presented in a recent review on time delay by de Carvalho and Nussenzveig \cite{CN}. In quantum-mechanical scattering theory time delay is relevant in particular for the characterization of resonances (\cite{Bohm}, \cite{Newton}, \cite{AS}), and it enters Levinson's theorem relating scattering data to the number of bound states (\cite{CN}, \cite{OB}). Time delay is related to the density of states in mesoscopic conductors (\cite{CN}, \cite{Buttiker}), to the virial coefficients in statistical mechanics (\cite{CN}, \cite{OT}), and it plays a role in the study of chaos (\cite{CN}, \cite{CHAOS}).

It is clear that the natural framework for defining the notion of time delay is that of time-dependent scattering theory. We refer to the book \cite{AJS} for a general account of this theory and just recall its basic ideas. One considers a physical system described by a Hamiltonian $H$ acting in some Hilbert space $\mathcal{H}$. The following questions are then considered: (i) Given a state vector $\psi \in \mathcal{H}$ at time $t=0$, what kind of asymptotic behavior can one expect to see for $\psi_{t} = e^{-iHt} \psi$ as $t \rightarrow \pm \infty$~? Often it is possible to show the absence of singular continuous spectrum of $H$, meaning roughly that - except for admitting superpositions - only two types of asymptotic behavior are possible: $\psi_{t}$ may stay essentially localized in a bounded region of configuration space (bound state) or disappear from each bounded region (scattering state). For state vectors of the second type one then asks: (ii) Can their evolution be described asymptotically (\ie for large values of $|t|$) in terms of a simpler Hamiltonian $H_{0}$ called the free Hamiltonian (or more generally in terms of a family of free Hamiltonians called channel Hamiltonians)~? The answer is positive in particular in situations where one can verify the existence of the M$\o$ller wave operators; this then leads to the introduction of the scattering operator $S$ that establishes the link between incoming and outgoing free states. A final important point then is: (iii) Can one describe the evolution of \emph{all} scattering states in terms of $H_{0}$ (or of the family of channel Hamiltonians)~? If so the theory is said to be asymptotically complete. Asymptotic completeness implies in particular the unitarity of $S$. Once the preceding dynamical questions have been settled one may introduce physical quantities like time delay or scattering cross sections and determine their properties and their relation to the $S$-operator. 

Very detailed results are known in time-dependent scattering theory for Hamiltonians describing a single non-relativistic particle in a potential $V(\mathbf{x})$ in $n$-dimensional space under the assumption that the potential tends to zero at large distances (\ie as $|\mathbf{x}| \rightarrow \infty$). On the other hand the literature on scattering theory in highly anisotropic situations, for example with potentials assuming different limits in different directions, is rather sparse (except for scattering relative to a periodic Hamiltonian and for one-dimensional Hamiltonians to be discussed below). For $n \geq 2$ the points (i)-(iii) have recently been investigated for potentials that are independent of $r \equiv |\mathbf{x}|$ outside some finite ball \cite{HS1} and for potentials with Cartesian anisotropy, \ie for potentials for which $\lim_{x_{j} \rightarrow \pm\infty} V(\mathbf{x})$ exist for each $j\in\{1,\dots,n\}$  \cite{Richard}. Typically a scattering system containing a highly anisotropic potential involves a multi-channel structure.

\begin{figure}[htbp]
  \begin{center}
       \scalebox{\scale}{\input{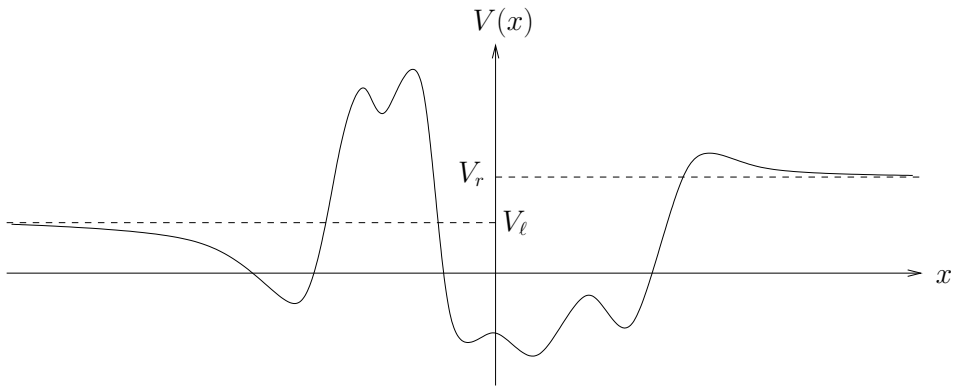}}\vspace{2mm}
       \caption{\normalsize A typical potential $V$.}\label{Fig1}
  \end{center}
\end{figure}

Time delay, for potentials vanishing at infinity, was first considered by Eisenbud~\cite{Eisenbud}, Bohm~\cite{Bohm} and Wigner~\cite{Wigner}. By using asymptotic properties of the solutions of the stationary Schr\"{o}dinger equation, they found that the energy derivative of the scattering phase shift may be interpreted as a time delay. Somewhat later Smith \cite{Smith} suggested, as we mentioned at the beginning, to consider the excess sojourn time $\tau_{\mathcal{X}}$ in a large spatial region $\mathcal{X}$ and to define time delay as the limit of $\tau_{\mathcal{X}}$ when this region tends to the entire configuration space $\mathbb{R}^n$. He showed, also in a stationary framework, that this leads again to the Eisenbud-Wigner expression. Smith's proposal was formalized in the framework of time-dependent scattering theory by Jauch and Marchand~\cite{JM}. These authors realized that the verification of the existence of the limit of  $\tau_{\mathcal{X}}$ as $\mathcal{X} \rightarrow \mathbb{R}^n$ represented a quite delicate mathematical problem. Later on a fair number of publications dealt with this problem; we refer to the review of Martin \cite{Martin} for details and references and mention that a satisfactory solution was given in \cite{AC}.
\newpage
In the present paper we shall consider the one-dimensional anisotropic case, \ie Hamiltonians of the form 
\begin{equation}\label{Hamiltonian}
H = H_{0} + V
\end{equation}
acting in the Hilbert space $\mathcal{H}=L^2(\mathbb{R})$. Here $H_{0} = P^2 = -d^2/dx^2$ is the usual free Hamiltonian and $V=V(Q)$ is given by a real-valued potential $V(x)$ assumed to have different limits at $x = -\infty$ and at $x = +\infty$; these limits will be denoted by $V_{\ell}$ and $V_r$ respectively  (see \fig\ref{Fig1}). Here we have set $\hbar=1$ for Planck's constant and $m=1/2$ for the mass of the particle, and we have written $P$ and $Q$ for the momentum and the position operator respectively in~$\mathcal{H}$. 

Hamiltonians of this type present a two-channel structure and can serve as models in the theory of mesoscopic quantum systems. Scattering theory for such one-dimensional Hamiltonians, with potentials having different limits on the left and on the right, has been investigated mostly in the time-independent formalism (\cite{CK}, \cite{Gesztesy}, \cite{GNP}). As regards the time-dependent approach,  the existence of the M$\o$ller wave operators was established in \cite{AK}, and a complete and detailed mathematical study of the questions (i)-(iii) mentioned before can be found in a paper by Davies and Simon \cite{DS}. As far as we know, time delay has been considered only for potentials for which $V_{\ell} = V_r$, \eg in \cite{JW1}; of course this special situation is also covered by our results.

We now describe the two representations of time delay in the above context, referring to Section~\ref{Time delay} for more details on our formalism. Because of the anisotropic structure of the potential there are two free Hamiltonians $H_{\ell}=H_{0}+V_{\ell}$ and $H_{r}=H_{0}+V_{r}$ entering into  the asymptotic description of the scattering states of $H$ and hence also into the definition of $S$. If $\psi \in \mathcal{H}$ is a scattering state of $H$, given by a normalized vector and interpreted as the state of a particle at time $t=0$, then its sojourn time (or dwell time) in the region $[-R,R]$ ($0 < R < \infty$) is given as follows: 
\begin{equation}\label{Sojourn-Time}
T_{R}(\psi) = \int_{-\infty}^{\infty} dt \int_{-R}^{R} dx \ |(e^{-iHt} \psi)(x)|^2\p.
\end{equation}

Similarly one can introduce free sojourn times with respect to $H_{\ell}$ and $H_{r}$. More specifically, let $\varphi \in \mathcal{H}$ be the initial state corresponding to $\psi$, \ie satisfying $\psi = \Omega^{-} \varphi$, where $\Omega^{-}$ is the M$\o$ller wave operator for the limit $t \rightarrow -\infty$. The incoming state $\varphi$ may be decomposed into a part incident from the left and a part incident from the right: $\varphi=\varphi_{\ell}+\varphi_{r}$. By using a similar decomposition of the associated outgoing state $S \varphi$ into a part $(S \varphi )_{\ell}$ propagating to the left and a part $(S \varphi )_{r}$ propagating to the right, one can then define the free incoming sojourn time $T^{\mathrm{in}}_{R}(\varphi)$ and the free outgoing sojourn time $T^{\mathrm{out}}_{R}(\varphi)$ associated to the initial state $\varphi$ by
\begin{eqnarray}
T^{\mathrm{in}}_{R}(\varphi) &=& \int_{-\infty}^{\infty} dt \int_{-R}^{R} dx \ |[e^{-iH_{\ell}t} \varphi_{\ell} + e^{-iH_{r}t} \varphi_{r}](x)|^2\p,\label{Sojourn-Time-In}\\
T^{\mathrm{out}}_{R}(\varphi) &=& \int_{-\infty}^{\infty} dt \int_{-R}^{R} dx \ |[e^{-iH_{\ell}t} (S \varphi )_{\ell} + e^{-iH_{r}t} (S \varphi )_{r}](x)|^2\p.\label{Sojourn-Time-Out}
\end{eqnarray}

The time delay in the interval $[-R,R]$ induced by the presence of the scatterer (represented by the potential $V$) is defined as the difference between the sojourn time of $\psi$ and the free sojourn times:
\begin{eqnarray}
\tau^{\mathrm{\mathrm{in}}}_{R}(\varphi) &=& T_{R}(\Omega^{-} \varphi) - T^{\mathrm{in}}_{R}(\varphi)\p,\label{Time-Delay-In}\\
\tau^{\mathrm{\mathrm{out}}}_{R}(\varphi) &=& T_{R}(\Omega^{-} \varphi) - T^{\mathrm{out}}_{R}(\varphi)\p.\label{Time-Delay-Out}
\end{eqnarray}

As will be seen in Section~\ref{Time delay}, these quantities are well defined for finite $R$. However for general states $\varphi\in\mathcal{H}$ they are divergent as $R\rightarrow\infty$, except when $V_{\ell} = V_{r}$ in which case they converge to the usual global time delay. The divergence when $V_{\ell} \not= V_r$ is not surprising: since the scattering is partially inelastic, the velocity in the state $e^{-iHt} \Omega^{-} \varphi_{\ell}$ may be different for example from that in the free state $e^{-iH_{\ell}t} \varphi_{\ell}$ (assuming the particle is incoming from the left) at large positive times, so that the local time delays $\tau^{\mathrm{in}}_{R}(\varphi)$ and  $\tau^{\mathrm{out}}_{R}(\varphi)$ will be proportional to $R$. A finite limit in this case, representing the \emph{global time delay} for the initial state $\varphi$, is obtained by starting from the following symmetrized expression for the local time delay:
\begin{equation}\label{Time-Delay-Average}
\tau_{R}(\varphi) = \frac{1}{2} \left[\tau^{\mathrm{in}}_{R}(\varphi) +
\tau^{\mathrm{out}}_{R}(\varphi)\right]\p.
\end{equation}

It is interesting to know that, even in anisotropic cases ($V_{\ell} \not= V_r$), the (symmetrized) global time delay is identical with the time-independent Eisenbud-Wigner expression of time delay which is often employed in calculations and introduced by a somewhat formal argument using the notion of group velocity  \cite{CN}. In particular, this identity permits one to have a time-dependent interpretation of the latter (see Section~\ref{Interpretation}). In terms of the $S$-matrix $S(E)$ at energy $E$, the Eisenbud-Wigner time delay operator at energy $E$ is given as
\begin{equation}\label{EW Matrix}
\mathcal{T}(E) = -i S(E)^{*} \frac{dS(E)}{dE}\p.
\end{equation}
The family $\{\mathcal{T}(E)\}$ determines a self-adjoint operator $\mathcal{T}$ in the Hilbert space $L^2(\mathbb{R})$, and the Eisenbud-Wigner time delay is defined as the expectation of this observable  $\mathcal{T}$ in the initial state $\varphi$:
\begin{equation}\label{EW Time Delay}
\tau^{\mathrm{EW}}(\varphi) \equiv \langle \varphi | \mathcal{T} \varphi \rangle =
\int_{-\infty}^{\infty} (\varphi(E) | \ \mathcal{T}(E) \varphi(E)) \, dE\p,
\end{equation}
where $\langle \cdot | \cdot \rangle$ denotes the scalar product in $L^2(\mathbb{R})$,  $( \cdot | \cdot )$ that at energy $E$ (details are given in Section~\ref{Global and Eisenbud-Wigner time delay}).

To end this introduction we present our assumptions on $V$ and outline the organization of this paper. We assume that the potential $V$ is given by a real-valued function satisfying  
\begin{eqnarray}
|V(x)-V_{\ell}| &\leq& M (1+|x|)^{-\mu} \hspace{3mm} \mbox{for} \ x \leq 0\p,\label{Assumption VL}\\
|V(x)-V_{r}| &\leq& M (1+|x|)^{-\mu} \hspace{3mm} \mbox{for} \  x
> 0\p,\label{Assumption VR}
\end{eqnarray}
for some positive constants $M$ and $\mu$. Here $V_{\ell}$ and $V_r$ are real numbers, with $V_{\ell} \leq V_r$. The constant $\mu$ specifies the rate at which $V(x)$ approaches its asymptotic limits. We assume throughout the paper that $\mu > 1$ (short range condition). Some results will be derived only under further conditions on $\mu$; the strongest hypothesis used is $\mu > 5$ (Section~\ref{Local time delay}). 

The next three sections are devoted to time-dependent scattering theory. Rather than citing results from the paper by Davies and Simon \cite{DS}, who
based their analysis on the Kato-Birman theory for trace class operators,
we give a presentation in the framework of the more recently developed
technique of differential inequalities (also called Mourre theory). This
method, described in Section~\ref{Time decay of wave packets}, will lead to various estimates on the rate
of decay of wave packets at large times and then, in Section~\ref{Large time limits}, to the characterization of
channel subspaces. In Section~\ref{Scattering theory} we discuss wave operators, asymptotic
completeness and the $S$-matrix $S(E)$. Finally, in Section~\ref{Time delay} we use
scattering theory to prove the existence of the limit defining the global
time delay and show its identity with the Eisenbud-Wigner representation. A
few technical points will be explained in the Appendices.

\section{Time decay of wave packets}\label{Time decay of wave packets}

Time-dependent scattering theory is based on properties of the
time evolution of wave packets and observables at large times $t$.
One has to know that such quantities have limits as $t \rightarrow
\pm \infty$ or that they decay sufficiently rapidly in time (\eg
the integrals in (\ref{Sojourn-Time})-(\ref{Sojourn-Time-Out})
defining sojourn times in bounded regions of configuration space
should be finite). We describe here some basic estimates on time
decay that will be used in the subsequent developments. The
derivation of these results will be given in the Appendices.

We write $H_{\kappa} = H_{0} + \kappa$
for the Hamiltonian with a constant potential given by the real number
$\kappa$ and denote by $\mathcal{F} \varphi$ or
$\hat{\varphi}$ the Fourier transform of a wave fonction $\varphi$:
\begin{equation}\label{Fourier Transform}
(\mathcal{F}\varphi)(p) \equiv \hat{\varphi}(p) = \frac{1}{\sqrt{2\pi}} \int_{-\infty}^{\infty} e^{-ipx} \varphi(x) \, dx\p.
\end{equation}

The spectral properties of $H$ and $H_{\kappa}$ are of course well
known. The spectrum of $H_{\kappa}$ is purely (absolutely)
continuous and covers the interval $[\kappa,+\infty)$. For the
class of potentials considered here, \ie satisfying
(\ref{Assumption VL})-(\ref{Assumption VR}) with $V_{\ell} \leq V_r$ and $\mu > 1$, the spectrum of $H$
consists of an (absolutely) continuous part coinciding with the
interval $[V_{\ell},+\infty)$ and possibly a set of non-degenerate eigenvalues $E_{k} \leq V_{\ell}$; if
$\mu > 2$, the number of eigenvalues is finite \cite{CCC}. We shall denote
by $\mathcal{H}_{p}(H)$ the subspace of the
Hilbert space $\mathcal{H}=L^2(\mathbb{R})$ spanned by the eigenvectors of the Hamiltonian $H$
and by $\mathcal{H}_{c}(H)$ its orthogonal complement in
$\mathcal{H}$, corresponding to the continuous spectrum of
$H$. The wave functions in $\mathcal{H}_{c}(H)$ are the scattering
states of the Hamiltonian $H$, also characterized by the property
that they disappear at large positive and negative times from each
bounded region in configuration space:
\begin{equation}\label{Scattering state}
\psi \in \mathcal{H}_{c}(H)  \Longleftrightarrow \int_{-R}^{R} \vert (e^{-iHt} \psi)(x) \vert^2 \, dx \rightarrow 0
\end{equation}
as $t \rightarrow \pm \infty$ for each fixed $R \in (0,\infty)$. For a Hamiltonian $H_{\kappa}$ with constant potential the subspace $\mathcal{H}_{c}(H_{\kappa})$
 is of course the entire space $L^{2}(\mathbb{R})$, and $\mathcal{H}_{p}(H_{\kappa})$ consists only of the zero element of $\mathcal{H}$,
\ie $\mathcal{H}_{p}(H_{\kappa}) = \{0\}$.

\subsection{Time decay in a constant potential}\label{Time decay in a constant potential}

In configuration space the time evolution of a wave packet
$\varphi$ in a constant potential is explicitly given as follows in terms of
the free propagator (Lemma 3.12 in \cite{AJS}):
\begin{equation}
\varphi_{t}(x) \equiv (e^{-iH_{\kappa}t}\varphi)(x) = \frac{e^{-i\kappa
t}}{\sqrt{4\pi i t}} \int_{-\infty}^{\infty} \exp\left(i
\frac{\vert x-y\vert^2}{4t}\right) \varphi(y) \, dy\p.
\end{equation}
From this expression one easily obtains the following formula for
the associated evolution in the Heisenberg picture of an
observable $f(Q)$, where $f$ is a function of a real variable:
\begin{equation}\label{Heisenberg Picture}
e^{iH_{\kappa}t} f(Q) e^{-iH_{\kappa}t} = e^{-iQ^2/4t} f(2tP) e^{iQ^2/4t}\p. 
\end{equation}
As $\exp(ix^{2}/4t)$ converges to $1$ when $t \rightarrow \pm
\infty$, this equation expresses the fact that, at large times,
position behaves approximately as momentum multiplied by $t/m$.

Since $\exp(iH_{\kappa}t)$ and $\exp(-iQ^{2}/4t)$ are unitary
operators, one obtains from (\ref{Heisenberg Picture}) the following important identity (by taking
into account the Plancherel theorem, \ie the unitarity of the Fourier
transformation $\mathcal{F}$):
\begin{equation}\label{Identity x-p}
\int_{-\infty}^{\infty} \vert f(x) \varphi_{t}(x)\vert^2 dx = \int_{-\infty}^{\infty} \vert f(2tp) \vert^2 \vert [\mathcal{F}(e^{iQ^2/4t} \varphi)](p)\vert^2 \, dp\p.
\end{equation}
This can be used for obtaining time decay, in the $L^{2}$-sense,
of quantities of the form $f(Q) e^{-iH_{\kappa}t}\varphi$ under
assumptions on the momentum distribution of $\varphi$. 

We shall
need the following estimates which can be obtained from (\ref{Identity x-p}) by taking for
$f$ a function of the Heaviside type or a function related to the potential V (see
Appendix~A). We say that $\varphi$ is a wave
packet with momentum in a subset $\Delta$ of the real line if
$\hat{\varphi}(p) = 0$ for all $p \not\in \Delta$. We are particularly
interested in decay properties at positive or negative times of
wave functions with positive or negative momentum or with no non-zero momentum components close to $p=0$. We have:\\ 
(a) Let $\varphi$ be a wave packet with positive momentum (\ie with momentum
in $(0,\infty)$), $x_{0}$ a real number and $\theta > 0$. Then there exists a constant $C_{\theta}$ so that for $t > 0$:
\begin{equation}\label{Propagation-a}
\int_{-\infty}^{x_{0}} \vert \varphi_{t}(x) \vert^2 \, dx \leq C_{\theta} (1+t)^{-\theta} \int_{-\infty}^{\infty} \vert (1+\vert x \vert)^{\theta} \varphi(x+x_{0}) \vert^2 \, dx
\end{equation}
and for $t < 0$:
\begin{equation}\label{Propagation-a2}
\int_{x_{0}}^{\infty} \vert \varphi_{t}(x) \vert^2 \, dx \leq C_{\theta} (1+\vert t \vert)^{-\theta} \int_{-\infty}^{\infty} \vert (1+\vert x \vert)^{\theta} \varphi(x+x_{0}) \vert^2 \, dx\p.
\end{equation}
(b) Let $\varphi$ be a wave packet with negative momentum (\ie
$\hat{\varphi}(p) = 0$ for $p\geq 0$), $x_{0} \in \mathbb{R}$ and $\theta > 0$. Then there exists a constant $C_{\theta}$ so that for $t > 0$:
\begin{equation}\label{Propagation-b}
\int_{x_{0}}^{\infty} \vert \varphi_{t}(x) \vert^2 \, dx \leq C_{\theta} (1 + t)^{-\theta} \int_{-\infty}^{\infty} \vert (1+\vert x \vert)^{\theta} \varphi(x+x_{0}) \vert^2 \, dx
\end{equation}
and for $t < 0$:
\begin{equation}\label{Propagation-b2}
\int_{-\infty}^{x_{0}} \vert \varphi_{t}(x) \vert^2 \, dx \leq C_{\theta} (1+\vert t \vert)^{-\theta} \int_{-\infty}^{\infty} \vert (1+\vert x \vert)^{\theta} \varphi(x+x_{0}) \vert^2 \, dx\p.
\end{equation}
(c) Let $f$ be a function satisfying $|f(x)| \leq C (1 + |x|)^{-\mu}$ for all $x \in\mathbb{R}$, some constant $C$ and some $\mu > 0$. Let $\varphi$ be a wave packet with momentum in $\mathbb{R}\setminus (-p_{0},p_{0})$ for some $p_{0} > 0$, and let $\theta > 0$. Then there exists a constant $C_{\theta}$ so that for all $t \in \mathbb{R}$:
\begin{equation}\label{Propagation-c}
\int_{-\infty}^{\infty} \vert f(x) \varphi_{t}(x) \vert^2 \, dx \leq C_{\theta} \left[\frac{1}{(1+\vert t \vert)^{\theta}} + \frac{1}{(1+2 p_{0} \vert t \vert)^{2\mu}}\right] \int_{-\infty}^{\infty} \vert (1+\vert x \vert)^{\theta} \varphi(x) \vert^2 \, dx\p.
\end{equation}

Of course the above estimates are useful only if the integrals on the \rhs are finite. This requirement essentially amounts to a differentiability property of the Fourier transform of $\varphi$ ($\hat{\varphi}$ should be $\theta$ times differentiable in some sense).

\subsection{Time decay in a non-constant potential}\label{Time decay in a non-constant potential}

Results on time decay of wave packets in a non-constant potential
are not so easy to obtain. The time evolution is now given by the
unitary operators $\exp(-iHt)$, and it is useful to relate them to the Green's
operator $(H-z)^{-1}$, with $z \in \mathbb{C}\setminus\mathbb{R}$. For $\epsilon > 0$ and
$E \in \mathbb{R}$ we set
\begin{equation}
 \delta_{(\epsilon)}(H-E) = \frac{1}{2\pi i} [(H-E-i\epsilon)^{-1} - (H-E+i\epsilon)^{-1}]\p.
\end{equation}
Then 
\begin{equation}\label{Delta(H-E)}
 \delta_{(\epsilon)}(H-E) = \frac{1}{2\pi} \int_{-\infty}^{\infty} e^{iEt} e^{-iHt - \epsilon \vert t \vert} \, dt\p.
\end{equation}
If $\psi$ is a (square-integrable) wave function, we set $\psi_{t}
= e^{-iHt}\psi$ and obtain from (\ref{Delta(H-E)}) by using the
Plancherel theorem that, for any bounded operator $B$:
\begin{equation}
\int_{-\infty}^{\infty} dt \int_{-\infty}^{\infty} dx \ e^{-2 \epsilon \vert t \vert} \vert(B \psi_{t})(x) \vert^{2} = 2\pi\int_{-\infty}^{\infty} dE \int_{-\infty}^{\infty} dx \ \vert [B \delta_{(\epsilon)}(H-E) \psi](x) \vert^{2}\p.
\end{equation}
Let us consider wave functions $\psi$ which have non-zero
components only in some finite closed interval $\Delta = [\alpha,\beta]$ in a
representation where the Hamiltonian $H$ is diagonal; such wave
functions $\psi$ will be said to have energy support in $\Delta$
(with respect to $H$). For such a wave function and for energies
$E$ not in $\Delta$, $\delta_{(\epsilon)}(H-E)\psi$ converges to
zero as $\epsilon \rightarrow 0$ (as $\delta_{(\epsilon)}(x)$
approximates the Dirac delta function), so that \cite{Comment-1} 
\begin{equation}\label{Limit-Delta}
\int_{-\infty}^{\infty} dt \int_{-\infty}^{\infty} dx \ \vert (B \psi_{t})(x) \vert^{2} = 2\pi \lim_{\epsilon \rightarrow 0} \int_{\alpha}^{\beta} dE \int_{-\infty}^{\infty} dx \ \vert [B \delta_{(\epsilon)}(H-E) \psi](x) \vert^{2}\p.
\end{equation}
If one knows that there is a constant $C$ such that
\begin{equation}\label{L2 bounds}
\int_{-\infty}^{\infty} \vert [B \delta_{(\epsilon)}(H-E)
\psi](x) \vert^{2} \, dx \leq C
 \end{equation}
for all energies $E$ in $\Delta$ and
all $\epsilon$ in an interval $(0,\epsilon_{0})$ for some $\epsilon_{0} > 0$,
then (\ref{Limit-Delta}) implies that the $L^{2}$-norm of
$B\psi$$_{t}$ is square-integrable over time, \ie
\begin{equation}\label{Propagation-Estimate}
\int_{-\infty}^{\infty} dt \int_{-\infty}^{\infty} dx \ \vert(B \psi_{t}
)(x) \vert^{2} \equiv \int_{-\infty}^{\infty} \| B \psi_{t}\|^2 \, dt \leq 2 \pi (\beta-\alpha) C < \infty\p.
\end{equation}
In (\ref{Propagation-Estimate}) and occasionally also further on we use the notation $\|\phi\|$ for the Hilbert space norm of a wave function $\phi$, \ie $\|\phi\| = [\langle\phi |\phi \rangle]^{1/2} = [\int_{-\infty}^{\infty} |\phi(x)|^2 \, dx]^{1/2}$.

Bounds on $L^{2}$-norms of the form (\ref{L2 bounds}) are a consequence of a
Mourre estimate. Roughly the validity of a Mourre estimate means that, for some suitable self-adjoint
operator $A$, the commutator between the Hamiltonian $H$ and $iA$ is strictly
positive on wave functions $\psi$ of the type used above (\ie having non-zero
energy components only in $\Delta$). More precisely, a Mourre estimate is
satisfied on an interval $\Delta$ if there are a self-adjoint operator $A$, a
compact operator $K$ and a real number $\lambda > 0$ such that, for each
$\psi$ having non-zero energy components only in $\Delta$, the following
inequality holds:
\begin{equation}\label{Mourre Estimate}
\langle \psi | [H,iA] \psi \rangle \geq \lambda \langle \psi | \psi \rangle + \langle \psi | K \psi \rangle\p.
\end{equation}
In general $A$ will be an unbounded operator, so that some
additional technical conditions must be imposed in order for the
commutator term occurring on the \lhs of (\ref{Mourre Estimate})
to be well defined. 

Mourre theory, based on an inequality of the type (\ref{Mourre Estimate}), is an abstract method for studying general self-adjoint operators $H$. For the Hamiltonians considered here (\ie one-dimensional Schr\"{o}dinger operators) a
suitable operator $A$ is given by $A = (PQ + QP)/4$. As shown in
Appendix~B, a Mourre estimate is then satisfied on each interval above $V_\ell$ disjoint from the scattering thresholds, more precisely on each interval $\Delta = [\alpha,\beta]$ with $V_{\ell} <  \alpha < \beta < V_{r}$ or $V_{r} < \alpha < \beta < \infty$.

We mention some interesting general consequences of a Mourre
estimate \cite{Mourre}. (i) In each interval $\Delta_{0} =
[\alpha_{0},\beta_{0}]$ in the interior of $\Delta$ (\ie $\alpha < \alpha_{0}<
\beta_{0}< \beta$), the Hamiltonian $H$ has at most a finite number of
eigenvalues (bound states), each at most finitely degenerate. (ii)
The continuous spectrum of $H$ in $\Delta$ is absolutely
continuous ($H$ has no Cantor-type spectrum in $\Delta$). (iii) If
$\Delta_{0} = [\alpha_{0},\beta_{0}]$ is an interval in the interior of
$\Delta$ and disjoint from the eigenvalues of $H$, then the norm
of the Green's operator $(H-E-i\epsilon)^{-1}$, sandwiched between
two operators $(1+|A|)^{-1}$, remains bounded near the real axis
(\ie for small values of $\epsilon$) at all energies $E$ in
$\Delta_{0}$:
\begin{equation}\label{Bound}
\sup_{E \in \Delta_{0}, \epsilon \not= 0} \|(1+|A|)^{-1}
(H-E-i\epsilon )^{-1} (1+|A|)^{-1}\|_{\mathcal{B}(\mathcal{H})} \leq C_{0} < \infty\p,
\end{equation}
where $\| \cdot \|_{\mathcal{B}(\mathcal{H})}$ is the operator norm in the space $\mathcal{B}(\mathcal{H})$ of bounded operators acting in $\mathcal{H}$. A consequence of (\ref{Bound}) is the validity of a strong version
of the propagation estimate (\ref{Propagation-Estimate}), with $B
= (1+|A|)^{-1}$ and wave packets $\psi$ having non-zero energy
components only in $\Delta_{0}$, namely (Proposition~7.1.1 in
\cite{ABG})
\begin{equation}\label{Propagation-Estimate-Strong}
\int_{-\infty}^{\infty} dt \int_{-\infty}^{\infty} dx \ \vert (B
\psi_{t})(x) \vert^{2} \leq 8 C_{0} \, \|\psi\|^2\p.
\end{equation}

In our situation the properties (i) and (ii) (even absence of
bound states in $(V_{\ell},\infty)$) were established independently of a
Mourre estimate, as already mentioned before. However property
(iii) is crucial for the approach to scattering theory presented
below. Indeed, if (\ref{Bound}) is satisfied with the operator $A$
indicated above, \ie $A = (PQ + QP)/4$, then it also holds
with $(1+|A|)^{-1}$ replaced by $(1+|Q|)^{-1}$ (with possibly some
different constant $C_{0}$ \cite{Comment-2}). This implies the
validity of the propagation estimate (\ref{Propagation-Estimate-Strong}) with $B = (1+|Q|)^{-1}$.
Consequently, if $f$ is a function of a real variable satisfying
$|f(x)| \leq C(1+|x|)^{-1}$ for some constant $C$ and all $x$,
then there is a constant $C_{1}$ such that for each $\psi$ with
energy support in $\Delta_{0}$:
\begin{equation}\label{Propagation-Estimate-Strong-Consequence}
\int_{-\infty}^{\infty} dt \int_{-\infty}^{\infty} dx \ \vert f(x)
\psi_{t}(x) \vert^{2} \equiv \int_{-\infty}^{\infty} \|
f(Q)\psi_{t}\|^2 \, dt \leq C_{1} \, \|\psi\|^2\p.
\end{equation}
In particular, for the Hamiltonians considered here, the time integral in
(\ref{Propagation-Estimate-Strong-Consequence}) is finite for each scattering state $\psi$ of $H$ with bounded energy support disjoint from the scattering
thresholds $V_{\ell}$ and $V_{r}$.

\newpage
\section{Large time limits}\label{Large time limits}

The bounds on certain time integrals obtained in the preceding section can be used
to prove the existence of limits, as $t \rightarrow \pm \infty$, of operators of the form $e^{ih_1 t} J e^{-ih_2 t}$,
where $J$ is a bounded operator and $h_{1}%
$, $h_{2}$ are Hamiltonians. It will be enough to consider
situations where $J = g(Q)$ is multiplication in
$L^{2}(\mathbb{R})$ by a function $g(x)$. We are particularly
interested in the cases $g \equiv 1$, $g = \chi_{\ell}$ and $g =
\chi_{r}$, where $\chi_{\ell}(x) = 1$ for $x\leq 0$, $\chi_{\ell}(x) =
0$ for $x > 0$ and $\chi_{r} = 1-\chi_{\ell}$. The functions $\chi_{\ell}$ and
$\chi_{r}$ represent localization on the left and on the right
respectively. For technical reasons we shall also
introduce smooth approximations of $\chi_{\ell}$ and $\chi_{r}%
$. We use the notation $\chi_{\ell}$ for the function $\chi_{\ell}$ and for the
operator $\chi_{\ell}(Q)$ of multiplication by this function.

All limits will be strong limits. We recall that, if $\{W_{t}\}_{t\in\mathbb{R}}$ and $W_{\infty}$ are bounded operators, then $\mbox{s-lim}_{t\rightarrow +\infty} W_{t}=W_{\infty}$ means that,
for each wave packet $\psi$, $W_{t}\psi - W_{\infty}\psi$
converges to zero in the Hilbert space norm, or equivalently that
$\{W_{t}\psi\}$ is Cauchy in the Hilbert space norm as $t \rightarrow
+\infty$. We first derive a useful formula (Eq. (\ref{W Cauchy}))
for verifying that $\{W_{t}\psi\}$ has this property. For $0 < s < t$  we have in the Hilbert space norm:
\begin{eqnarray}
\|W_{t}\psi - W_{s}\psi\| &=& \sup_{\|\phi\|=1} |\langle \phi | (W_{t} - W_{s}) \psi\rangle|\nonumber\\
&=& \sup_{\|\phi\|=1} \left| \int_{s}^{t} [\frac{d}{d\tau}
\langle \phi | W_{\tau} \psi\rangle] \, d\tau\right|\p.
\end{eqnarray}
If $W_{\tau} = e^{ih_{1}\tau} J e^{-ih_{2}\tau}$ with
$h_{1}=-d^2/dx^2 + V_{1}(x)$,  $h_{2}=-d^2/dx^2 + V_{2}(x)$ and
$J=g(Q)$, then
\begin{equation}
\frac{d}{d\tau}W_{\tau} = i e^{ih_{1}\tau} (h_{1}J-Jh_{2}) e^{-ih_{2}\tau}
\end{equation}
and
\begin{eqnarray}
h_{1}J-Jh_{2} &=& -g'' - 2g'\frac{d}{dx} + (V_{1}-V_{2})g\nonumber\\
&=& -g''(Q) -2ig'(Q)P + (V_{1}-V_{2})g(Q)\p.
\end{eqnarray}
Hence
\begin{equation}\label{W Cauchy}
\|W_{t}\psi -W_{s}\psi\| \leq \sup_{\|\phi\|=1}
\int_{s}^{t} |\langle e^{-ih_{1}\tau} \phi | [-g''(Q) -2ig'(Q)P +
(V_{1}-V_{2})g(Q)] e^{-ih_{2}\tau}\psi\rangle| \, d\tau\p.
\end{equation}

The inequality (\ref{W Cauchy}) allows one to infer that $\{W_{t}\psi\}$ is Cauchy as $t \rightarrow \pm\infty$ if a suitable estimate on the time decay of the integrand is available. If $V_2$ is a constant potential, (\ref{W Cauchy}) corresponds to the well-known Cook method for proving the existence of limits of the type considered here. In this case it suffices to know the simple decay estimates of Section~\ref{Time decay in a constant potential}; an application concerns the existence of the M$\o$ller wave operators in Section~\ref{Existence and properties of the channel wave and scattering operators}. More refined techniques are needed for estimating the integral in  (\ref{W Cauchy}) if $V_2$ is a non-constant potential. Using estimates of the type (\ref{Propagation-Estimate-Strong-Consequence}), deduced from Mourre theory, we shall obtain two important results for scattering theory (existence of the limits in (\ref{Projectors-L}), (\ref{Projectors-R}) and (\ref{Separating LR3})); details are presented in Appendix~C.

\subsection{The channel subspaces}\label{The free channel subspaces}

In the simple case where $h_{1} = h_{2} = H_{\kappa}$ and $J =
\chi_{\ell}$, convergence can be obtained without making use of an
estimate of the form  (\ref{W Cauchy}). To know that $e^{iH_\kappa
t} \chi_{\ell} e^{-iH_\kappa t}$ converges strongly as $t \rightarrow
+\infty$, it suffices to show that $\mbox{s-lim}_{t \rightarrow
+\infty} e^{iH_\kappa t} \chi_{\ell} e^{-iH_\kappa t} \varphi$ exists
for a dense set of wave packets $\varphi$. We consider the
following dense set: $\varphi = \varphi_{+} + \varphi _{-}$,
where $\varphi_{+}$ has positive momentum, $\varphi_{-}$ has
negative momentum and $(1+|Q|)^{\theta}\varphi_{\pm}$ are
square-integrable for some $\theta > 0$. The limit as $t \rightarrow +\infty$ of
$\chi_{\ell} e^{-iH_\kappa t} \varphi_{+}$ is zero,
which expresses the fact that $e^{-iH_\kappa t}\varphi_{+}$
propagates towards the right (take $x_{0} = 0$ and $\theta > 0$ in
(\ref{Propagation-a})). Similarly (\ref{Propagation-b}) implies
that $\mbox{s-lim}_{t \rightarrow +\infty} \chi_{r} e^{-iH_\kappa t} \varphi_{-}=0$. Since $e^{iH_{\kappa}t}$ is unitary, it follows that $e^{iH_\kappa
t}\chi _{\ell} e^{-iH_\kappa t} \varphi_{-} = \varphi_{-} -
e^{iH_\kappa t} \chi_{r} e^{-iH_\kappa t} \varphi_{-}$ converges
to $\varphi_{-}$ as $t \rightarrow + \infty$. In conclusion:
$\mbox{s-lim}_{t \rightarrow +\infty} e^{iH_\kappa t} \chi_{\ell}
e^{-iH_\kappa t} \varphi = \varphi_{-}$.

As an operator, $\mbox{s-lim}_{t\rightarrow +\infty} e^{iH_\kappa
t} \chi_{\ell} e^{-iH_\kappa t}$ represents the observable of
localization on the left at $t = +\infty$ in a constant
potential; it does not depend on $\kappa$ and will be denoted
by $F^{+}_{0,\ell}$. So  $F^{+}_{0,\ell}$ is the (orthogonal) projection
onto the subspace $\mathcal{H}^{+}_{0,\ell}$ of wave functions that are
localized on the left at $t = +\infty$ (in a constant potential), and it
coincides with the projection $\Pi_{-}$ onto the subspace
$\mathcal{H}_{-}$ of wave functions with negative momentum.

One can similarly obtain the existence of the following limits and
relate them to $\Pi_{-}$ or to the projection $\Pi_{+}$ onto the
subspace $\mathcal{H}_{+}$ of wave packets with positive momentum:
\begin{eqnarray}
F^{+}_{0,\ell} &=& \begin{array}{c}
  \vspace{\slim} \mbox{s-lim} \\
   \vspace{\tlim} \mbox{\scriptsize $t$ $\rightarrow$ $+\infty$}
\end{array} \ e^{iH_\kappa t} \chi_{\ell} e^{-iH_\kappa t} = \Pi_{-}\p,\label{Projectors-PL}\\
F^{-}_{0,\ell} &=& \begin{array}{c}
  \vspace{\slim} \mbox{s-lim} \\
   \vspace{\tlim} \mbox{\scriptsize $t$ $\rightarrow$ $-\infty$}
\end{array} \ e^{iH_\kappa t} \chi_{\ell} e^{-iH_\kappa t} = \Pi_{+}\p,\label{Projectors-ML}\\
F^{+}_{0,r} &=& \begin{array}{c}
  \vspace{\slim} \mbox{s-lim} \\
   \vspace{\tlim} \mbox{\scriptsize $t$ $\rightarrow$ $+\infty$}
\end{array} \ e^{iH_\kappa t} \chi_{r} e^{-iH_\kappa t} = \Pi_{+}\p,\label{Projectors-PR}\\
F^{-}_{0,r} &=& \begin{array}{c}
  \vspace{\slim} \mbox{s-lim} \\
   \vspace{\tlim} \mbox{\scriptsize $t$ $\rightarrow$ $-\infty$}
\end{array} \ e^{iH_\kappa t}
\chi_{r} e^{-iH_\kappa t} = \Pi_{-}\p.\label{Projectors-MR}
\end{eqnarray}
In terms of these operators, the propagation properties of $e^{-iH_{\kappa}t} \varphi_{\pm}$ pointed out above may be expressed as follow:
\begin{equation}\label{Orthogonality Fpm}
\begin{array}{c}
  \vspace{\slim} \mbox{s-lim} \\
   \vspace{\tlim} \mbox{\scriptsize $t$ $\rightarrow$ $+\infty$}
\end{array} \ \chi_{\ell} e^{-iH_\kappa t} F^{+}_{0,r} =  \begin{array}{c}
  \vspace{\slim} \mbox{s-lim} \\
   \vspace{\tlim} \mbox{\scriptsize $t$ $\rightarrow$ $+\infty$}
\end{array} \ \chi_{r} e^{-iH_\kappa t} F^{+}_{0,\ell} = 0\p.
\end{equation}

\subsection{Subspaces of scattering states}\label{The total channel subspaces}

In the presence of a potential it is also possible to divide the
set of scattering states $\mathcal{H}_{c}(H)$ into two mutually
orthogonal subspaces $\mathcal{H}^{+}_{\ell}$ and
$\mathcal{H}^{+}_{r}$ containing the state vectors localized on
the left and on the right respectively at $t=+\infty$. For this
one shows that the strong limits of $e^{iHt}\chi_{\ell}e^{-iHt}$ and
$e^{iHt}\chi_{r}e^{-iHt}$ as $t \rightarrow +\infty$ exist on
$\mathcal{H}_{c}(H)$ and define two projections $F^{+}_{\ell}$ and
$F^{+}_{r}$ in $\mathcal{H}_{c}(H)$, with  $F^{+}_{\ell} F^{+}_{r} =
0$. A similar decomposition $\mathcal{H}_{c}(H)
=\mathcal{H}^{-}_{\ell} \oplus \mathcal{H}^{-}_{r}$ exists,
corresponding to the limit $t \rightarrow -\infty$. In contrast to the case of a constant potential (where $\mathcal{H}^{-}_{0,\ell}=\mathcal{H}^{+}_{0,r}$), these two decompositions of $\mathcal{H}_{c}(H)$ are different in general; indeed the equality of $\mathcal{H}^{-}_{\ell}$ and $\mathcal{H}^{+}_{r}$ would mean that the potential is reflectionless. On the other hand $\mathcal{H}^{+}_{\ell}$ and $\mathcal{H}^{-}_{\ell}$ (and similarly  $\mathcal{H}^{+}_{r}$ and $\mathcal{H}^{-}_{r}$) are related by time reversal. The antiunitary time reversal operator $\Theta$, given by complex conjugation, \ie $(\Theta \psi)(x) = \overline{\psi(x)}$, commutes with the Hamiltonian $H$ and with $\chi_{\ell}$, so that $\Theta e^{iHt} \chi_{\ell} e^{-iHt}\psi = e^{-iHt} \chi_{\ell} e^{iHt} \Theta\psi$. It follows that $\psi$ belongs to $\mathcal{H}^{+}_{\ell}$ if and only if $\Theta\psi$ belongs to $\mathcal{H}^{-}_{\ell}$. Obviously these relations also hold for $\mathcal{H}^{\pm}_{0,\ell}$ and $\mathcal{H}^{\pm}_{0,r}$, \ie $\Theta \mathcal{H}^{\pm}_{0,\ell} = \mathcal{H}^{\mp}_{0,\ell}$ and $\Theta \mathcal{H}^{\pm}_{0,r} = \mathcal{H}^{\mp}_{0,r}$. (An equation bearing double signs always has to be interpreted as two independent equations, one for the upper and one for the lower sign).

Denoting the projection onto the subspace $\mathcal{H}_{c}(H)$ of scattering states by $F_{c}(H)$, we set
\begin{eqnarray}
F^{\pm}_{\ell} &=& \begin{array}{c}
  \vspace{\slim} \mbox{s-lim} \\
   \vspace{\tlim} \mbox{\scriptsize $t$ $\rightarrow$ $\pm$ $\infty$}
\end{array} \ e^{iHt}\chi_{\ell} e^{-iHt}F_{c}(H)\p,\label{Projectors-L}\\
F^{\pm}_{r} &=& \begin{array}{c}
  \vspace{\slim} \mbox{s-lim} \\
   \vspace{\tlim} \mbox{\scriptsize $t$ $\rightarrow$ $\pm$ $\infty$}
\end{array} \ e^{iHt}\chi_{r} e^{-iHt}F_{c}(H)\p.\label{Projectors-R}
\end{eqnarray}
The existence of these limits can be obtained by using (\ref{W Cauchy}) with $W_{\tau}=e^{iH\tau} g(Q) e^{-iH\tau}$, where $g$ is a smooth approximation of $\chi_\ell$ or $\chi_r$. Details on this are given in Appendix~C. Below we deduce alternative expressions for these limits (Eqs. (\ref{F Existence}) and (\ref{F+L})) and determine their properties. We shall consider $F^{+}_{\ell}$.

Let $F_{p}(H) = \sum_{k=1}^{N} | \varphi_{k}\rangle\langle
\varphi_{k}|$ be the projection onto the subspace
$\mathcal{H}_{p}(H)$ of bound states of $H$: $\varphi_{1},\dots,
\varphi_{N}$ are normalized eigenfunctions if $H$ has $N$ bound
states (we consider the case where $N\not=0$). Denoting the
eigenvalues by $E_{1},\dots, E_{N}$ we have
$e^{-iHt}\varphi_{k}=e^{-iE_{k}t}\varphi_{k}$, and the square of
the norm of $F_{p}(H)e^{iHt}\chi_{\ell}e^{-iHt}\psi$ is just
\begin{equation}
\sum_{k=1}^{N} |\langle
\varphi_{k}|e^{iHt}\chi_{\ell}e^{-iHt}\psi\rangle|^2 =
\sum_{k=1}^{N} |\langle
\varphi_{k}|\chi_{\ell}e^{-iHt}\psi\rangle|^2\p.
\end{equation}
We set $\psi_{t}=e^{-iHt}\psi$ and define the function
$\chi_{(\alpha,\beta]}$ by $\chi_{(\alpha,\beta]}(x)=1$ if $\alpha< x \leq \beta$ and
$\chi_{(\alpha,\beta]}(x)=0$ otherwise. Then $\chi_{\ell}\equiv
\chi_{(-\infty,0]}= \chi_{(-\infty,-R]} +  \chi_{(-R,0]}$ for any
$R > 0$, hence
\begin{equation}\label{Decomposition}
\langle \varphi_{k}|\chi_{\ell}\psi_{t}\rangle = \langle \chi_{(-\infty,-R]}\varphi_{k}|\psi_{t}\rangle + \langle \varphi_{k}|\chi_{(-R,0]}\psi_{t}\rangle\p.
\end{equation}
By the Schwarz inequality, the absolute square of the first term on the
\rhs of (\ref{Decomposition}) is bounded by $\int_{-\infty}^{-R}
|\varphi_{k}(x)|^2 \, dx \int_{-\infty}^{\infty} |\psi(y)|^2 \, dy$,
which can be made arbitrarily small by choosing $R$ large enough.
The absolute square of the second term is bounded by
$\int_{-\infty}^{\infty} |\varphi_{k}(x)|^2 \, dx \int_{-R}^{0}
|\psi_{t}(y)|^2 \, dy$, which converges to zero as $t \rightarrow
+\infty$ by (\ref{Scattering state}) if $\psi \in
\mathcal{H}_{c}(H)$. If $N$ is finite, we conclude that
$F_{p}(H)e^{iHt}\chi_{\ell}e^{-iHt}F_{c}(H)$ converges strongly to
zero as $t \rightarrow
+\infty$ \cite{Q}. Since $F_{p}(H)+F_{c}(H)=1$, it follows that the limit
defining $F^{+}_{\ell}$ exists if and only if
$\mbox{s-lim}_{t\rightarrow +\infty} F_{c}(H)
e^{iHt}\chi_{\ell}e^{-iHt}F_{c}(H)$ exists, and that
\begin{equation}\label{F Existence}
F^{+}_{\ell} = F_{c}(H) F^{+}_{\ell} = \begin{array}{c}
  \vspace{\slim} \mbox{s-lim} \\
   \vspace{\tlim} \mbox{\scriptsize $t$ $\rightarrow$ $+\infty$}
\end{array} \ F_{c}(H) e^{iHt}\chi_{\ell}e^{-iHt}F_{c}(H)\p.
\end{equation}

The step function $\chi_{\ell}$ in (\ref{F Existence}) may be replaced by a smooth approximation $g$. Let $g$ be a smooth function satisfying $0 \leq g(x) \leq 1$ for
all $x$, $g(x)=1$ for $x < -1$ and $g(x)=0$ for $x>0$. Then, by
(\ref{Scattering state}) with $R=1$,
$\chi_{\ell}e^{iHt}\psi-g(Q)e^{-iHt}\psi$ converges to zero as
$t\rightarrow +\infty$ for each $\psi\in\mathcal{H}_{c}(H)$, hence
\begin{equation}\label{F+L}
F^{+}_{\ell} = \begin{array}{c}
  \vspace{\slim} \mbox{s-lim} \\
   \vspace{\tlim} \mbox{\scriptsize $t$ $\rightarrow$ $+\infty$}
\end{array} \ F_{c}(H) e^{iHt} g(Q) e^{-iHt} F_{c}(H)\p.
\end{equation}

Let us mention some simple consequences of (\ref{F Existence}): (i) $F^{+}_{\ell}$ is self-adjoint, \ie $(F^{+}_{\ell})^{*}=F_{\ell}^{+}$. (ii) $F^{+}_{\ell}$ is idempotent. Indeed, using (i), the unitarity of $e^{iHt}$, the fact that $\chi_\ell^2 = \chi_\ell$ and the identity (\ref{F Existence}), one obtains
\begin{eqnarray}\label{F2=F}
\langle \varphi | (F^{+}_{\ell})^{2} \psi \rangle &=& \langle F^{+}_{\ell} \varphi | F^{+}_{\ell} \psi \rangle \nonumber\\
&=& \lim_{t \rightarrow \infty} \langle e^{iHt} \chi_\ell e^{-iHt} F_c(H) \varphi |  e^{iHt} \chi_\ell e^{-iHt} F_c(H) \psi \rangle\nonumber \\
&=& \lim_{t \rightarrow \infty} \langle \varphi | F_c(H) e^{iHt} \chi_\ell e^{-iHt} F_c(H) \psi \rangle\nonumber \\
&=& \langle \varphi | F^{+}_{\ell} \psi \rangle\p.
\end{eqnarray}
Hence $(F^+_\ell)^2=F_\ell^+$. The properties (i) and (ii) mean that $F^{+}_{\ell}$
is an orthogonal projection in $\mathcal{H}_{c}(H)$. The subspace
onto which it projects is denoted by $\mathcal{H}_{\ell}^{+}$. This
subspace is invariant under the evolution: $e^{-iHs}$ leaves
$\mathcal{H}_{\ell}^{+}$ invariant ($s \in \mathbb{R}$), or
equivalently $F^{+}_{\ell}$ commutes with $e^{-iHs}$ (and hence with
$H$). Indeed, since $e^{-iHs}$ commutes with $F_{c}(H)$:
\begin{eqnarray}
F^{+}_{\ell} e^{-iHs} &=& \begin{array}{c}
  \vspace{\slim} \mbox{s-lim} \\
   \vspace{\tlim} \mbox{\scriptsize $t$ $\rightarrow$ $+\infty$}
\end{array} \ e^{iHt} \chi_{\ell} e^{-iH(t+s)} F_{c}(H)\nonumber \\
&=& \begin{array}{c}
  \vspace{\slim} \mbox{s-lim} \\
   \vspace{\tlim} \mbox{\scriptsize $\tau$ $\rightarrow$ $+\infty$}
\end{array} \ e^{iH(\tau-s)} \chi_{\ell} e^{-iH \tau} F_{c}(H)\nonumber\\
 &=& e^{-iHs} F^{+}_{\ell}\p.
\end{eqnarray}
Also, by an argument as in (\ref{F2=F}), using the relation
$\chi_{r}\chi_{\ell}=0$, one finds that $F^{+}_{r} F^{+}_{\ell}=0$ or
\begin{equation}\label{Orthogonality Fpm2}
\begin{array}{c}
  \vspace{\slim} \mbox{s-lim} \\
   \vspace{\tlim} \mbox{\scriptsize $t$ $\rightarrow$ $+\infty$}
\end{array} \ \chi_{r} e^{-iH t} F^{+}_{\ell} = \begin{array}{c}
  \vspace{\slim} \mbox{s-lim} \\
   \vspace{\tlim} \mbox{\scriptsize $t$ $\rightarrow$ $+\infty$}
\end{array} \ \chi_{\ell} e^{-iH t} F^{+}_{r} = 0\p.
\end{equation}
This shows that $\mathcal{H}^{+}_{r}$ is orthogonal to
$\mathcal{H}^{+}_{\ell}$, and clearly $\mathcal{H}_{c}(H) =
\mathcal{H}^{+}_{r} \oplus \mathcal{H}^{+}_{\ell}$ since $F^{+}_{\ell} +
F^{+}_{r}=F_{c}(H)$.

\section{Scattering theory}\label{Scattering theory}

As explained in the Introduction one needs to answer some dynamical questions (points (i) to (iii)) in the framework of scattering theory in order to define rigorously the notion of time delay. We first introduce the asymptotic condition representing the fundamental idea of scattering theory. This leads us naturally to the introduction of the M$\o$ller wave operators $\Omega^{\pm}$ and then the scattering operator $S$. The anisotropic structure of the systems considered here is reflected in the appearance of channel operators. Existence, properties and the interpretation of these channel operators are then discussed. Finally we introduce the $S$-matrix $S(E)$ entering the Eisenbud-Wigner representation of time delay. 

\subsection{Asymptotic condition}\label{Asymptotic condition}

The fundamental idea of scattering theory, expressed in the context of anisotropic systems considered in this paper, is that at large (negative and positive) times $t$ a particle in a scattering state $\psi_{t} = e^{-iHt} \psi \in \mathcal{H}_{c}(H)$ is located in a region far from the scatterers, where the potential $V$ is essentially constant (approaching $V_{\ell}$ at $x=-\infty$ and $V_{r}$ at $x=+\infty$), and therefore should behave essentially as a free particle (evolving with $H_{\ell} = H_{0} + V_{\ell}$ and $H_{r} = H_0 + V_{r}$ respectively).

This idea is called \emph{the asymptotic condition} and is formalized as follows. Let $\psi \in \mathcal{H}_{c}(H)$ be a scattering state of $H$. Then there should exist free scattering states $\varphi^{\pm}_{\ell} \in \mathcal{H}$ of $H_{\ell}$ and $\varphi^{\pm}_{r} \in \mathcal{H}$ of $H_{r}$ so that
\begin{equation}\label{Asymptotic-Condition}
\lim_{t \rightarrow \pm \infty} \int_{\mathbb{R}} |(e^{-iHt} \psi)(x) - (e^{-iH_{\ell}t} \varphi^{\pm}_{\ell})(x) - (e^{-iH_{r}t} \varphi^{\pm}_{r})(x)|^2 \, dx = 0\p,
\end{equation}
where as before this condition has to be interpreted as two independent relations, one for each sign in $\pm$. 

\subsection{Wave and scattering operators}\label{Wave and scattering operators}

To show that the conditions (\ref{Asymptotic-Condition}) are satisfied,
we introduce the \emph{M$\o$ller wave operators}
\begin{equation}\label{Wave}
\Omega^{\pm} = \Omega^{\pm}_{\ell} + \Omega^{\pm}_{r}
\end{equation}
with
\begin{eqnarray}
\Omega_{\ell}^{\pm} &=& \begin{array}{c}
  \vspace{\slim} \mbox{s-lim} \\
   \vspace{\tlim} \mbox{\scriptsize $t$ $\rightarrow$ $\pm$ $\infty$}
\end{array} \ e^{iHt} e^{-iH_{\ell}t} F^{\pm}_{0,\ell}\p,\label{Channel-Wave L}\\
\Omega_{r}^{\pm} &=& \begin{array}{c}
  \vspace{\slim} \mbox{s-lim} \\
   \vspace{\tlim} \mbox{\scriptsize $t$ $\rightarrow$ $\pm$ $\infty$}
\end{array} \ e^{iHt} e^{-iH_{r}t} F^{\pm}_{0,r}\p,\label{Channel-Wave R}
\end{eqnarray}
where the projections $F^{\pm}_{0,\ell}$ and $F^{\pm}_{0,r}$ are those defined in (\ref{Projectors-PL})-(\ref{Projectors-MR}). We call the operators $\Omega_{\ell}^{\pm}$ and $\Omega_{r}^{\pm}$ the \emph{channel wave operators}. 

It is straightforward to see that the asymptotic condition (\ref{Asymptotic-Condition}) is equivalent to the existence of the wave operators $\Omega^{\pm}$ and their completeness, \ie their ranges $\mathcal{R}(\Omega^{\pm})$ are the entire continuous subspace $\mathcal{H}_{c}(H)$, justifying a posteriori the introduction of the projections $F^{\pm}_{0,\ell}$ and $F^{\pm}_{0,r}$ selecting the states in $\mathcal{H}$ which have the appropriate asymptotic behavior. These facts will be proved in the next subsection. Assuming this one can define the scattering operator by
\begin{equation}
S = (\Omega^{+})^{*} \Omega^{-}\p.
\end{equation}
The decomposition (\ref{Wave}) of the wave operators into a left and a right part leads to the following structure of $S$:
\begin{equation}
S = S_{\ell} + S_{r} = S_{\ell \ell} + S_{r \ell} + S_{rr} + S_{\ell r}\p,
\end{equation}
with $S_{\ell}=S_{\ell \ell} + S_{r \ell}$, $S_{r}= S_{rr} + S_{\ell r}$ and $S_{ab}=(\Omega_{a}^{+})^{*} \Omega_{b}^{-}$, where $a$ and $b$ stand for $\ell$ or $r$. We call the operators $S_{ab}$ the \emph{channel scattering operators}. Their properties and interpretation are established in the next subsection. In the sequel we shall use the subscripts $a$, $b$, $c$ and $d$ to denote $\ell$ or $r$. 

The introduction of the sum of the channel wave operators, Eq. (\ref{Wave}), may seem somewhat unusual for a multichannel system. In our context this sum represents an interesting and useful operator, as shown under point (iii) of the next subsection. It should be noted that the channel structure arising here is different from that occuring for example in the quantum-mechanical $N$-body problem. In the latter the channel subspaces may overlap but are independent of the sign of time, whereas for the scattering systems considered here the channel subspaces depend on the sign of time (see (\ref{Projectors-PL})-(\ref{Projectors-MR})) but represent, for each sign of time, a decomposition of the Hilbert space $\mathcal{H}$ into mutually orthogonal subspaces. The wave operators $\Omega^{\pm}$ in (\ref{Wave}) may be expressed in terms of the free Hamiltonians $H^{\mathrm{in}}$ and $H^{\mathrm{out}}$ introduced in (\ref{Free Hamiltonian}) as follows: $\Omega^{-}=\mbox{s-lim}_{t\rightarrow -\infty} e^{iHt} e^{-iH^{\mathrm{in}}t}$ and $\Omega^{+}=\mbox{s-lim}_{t\rightarrow +\infty} e^{iHt} e^{-iH^{\mathrm{out}}t}$. [The quantum-mechanical $N$-body problem is often written in a similar form, called the two-Hilbert space formulation (see \eg \cite{AJS} or \cite{BW}) by introducing an auxiliary asymptotic Hilbert space $\mathcal{H}^{\mathrm{as}}$ (the orthogonal direct sum of the channel subspaces) and an asymptotic free Hamiltonian $H^{\mathrm{as}}$ acting in $\mathcal{H}^{\mathrm{as}}$ and independent of the sign of time.]

\subsection{Existence and properties of the channel wave and scattering operators}\label{Existence and properties of the channel wave and scattering operators}

In this subsection we show that, if one assumes the potential $V$ to satisfy (\ref{Assumption VL})-(\ref{Assumption VR}) with $\mu > 1$, then the channel wave operators $\Omega^{\pm}_{a}$ are well defined, \ie the time limits involved in their definition (\ref{Channel-Wave L})-(\ref{Channel-Wave R}) exist. We then establish their properties and those of the channel scattering operators $S_{ab}$ (assuming that $\mu \geq 2$).

The existence of the channel wave operators can be obtained by invoking only properties of time decay in a constant potential. We show the existence of $\Omega^{+}_{\ell}$, the other channel wave operators can be handled similarly. Let $\Omega(t) =  e^{iHt} e^{-iH_{\ell}t} F^{+}_{0,\ell}$ and write
\begin{eqnarray}\label{Separating LR}
\Omega(t) =  e^{iHt} \chi_{\ell} e^{-iH_{\ell}t} F^{+}_{0,\ell} + e^{iHt} \chi_{r} e^{-iH_{\ell}t} F^{+}_{0,\ell}\p,
\end{eqnarray}
where $\chi_{\ell}$ and
$\chi_{r}$ represent the configuration space localization on the left and on the right
respectively introduced in Section~\ref{Large time limits}.

Using the unitarity of $e^{iHt}$ and (\ref{Orthogonality Fpm}) it follows that the second term on the \rhs of (\ref{Separating LR}) vanishes as $t \rightarrow +\infty$, so we have 
\begin{eqnarray}\label{Identity Lg}
\Omega^{+}_{\ell} &=& \begin{array}{c}
  \vspace{\slim} \mbox{s-lim} \\
   \vspace{\tlim} \mbox{\scriptsize $t$ $\rightarrow$ $+$ $\infty$}
\end{array} \ e^{iHt} \chi_{\ell} e^{-iH_{\ell}t} F^{+}_{0,\ell} \nonumber\\
 &=& \begin{array}{c}
  \vspace{\slim} \mbox{s-lim} \\
   \vspace{\tlim} \mbox{\scriptsize $t$ $\rightarrow$ $+$ $\infty$}
\end{array} \ e^{iHt} g(Q) e^{-iH_{\ell}t} F^{+}_{0,\ell}\p,
\end{eqnarray}
where $g$ is the smooth approximation of $\chi_{\ell}$ introduced before Eq. (\ref{F+L}).

We consider the following dense set of wave packets $\varphi$ in the subspace $\mathcal{H}_{0,\ell}^{+}$: $\varphi$ has (negative) momentum in a bounded closed set not containing $p=0$ and is such that $\hat{\varphi}$ is three times continuously differentiable. From (\ref{W Cauchy}) with $h_1 = H$ and $h_2 = H_{\ell}$, hence $W_{\tau} = e^{iH\tau} g(Q) e^{-iH_{\ell}\tau}$, one obtains that
\begin{equation}\label{W Exists2}
\|W_{t}\varphi - W_{s}\varphi\| \leq \int_{s}^{t} N_{\tau} \, d\tau\p,
\end{equation}
where
\begin{equation}\label{Bound2}
N_{\tau} = \|[-g''(Q) - 2i g'(Q)P + (V-V_{\ell})g(Q)]  e^{-iH_{\ell}\tau} \varphi\|\p.
\end{equation}
Since $g'$ and $g''$ vanish outside the interval  $[-1,0]$ and $|[V(x)-V_{\ell}] g(x)| \leq M (1 + |x|)^{-\mu}$ with $\mu > 1$, the integral $\int_{0}^{\infty} N_{\tau} \, d\tau$ is finite as a consequence of (\ref{Propagation-c}) (with $\theta = 3$). Thus $\Omega^{+}_{\ell}$ exists.

Next let $\mathcal{W}(t) =  e^{iH_{\ell} t} e^{-iH t} F^{+}_{\ell}$. As shown in Appendix~C, these operators are strongly convergent as $t \rightarrow + \infty$. By arguing as in (\ref{Separating LR})-(\ref{Identity Lg}) and using (\ref{Orthogonality Fpm2}) one obtains the following expression for their limit:
\begin{eqnarray}\label{Separating LR3}
\mathcal{W} \equiv \begin{array}{c}
  \vspace{\slim} \mbox{s-lim} \\
   \vspace{\tlim} \mbox{\scriptsize $t$ $\rightarrow$ $+$ $\infty$}
\end{array} \ \mathcal{W}(t) =  \begin{array}{c}
  \vspace{\slim} \mbox{s-lim} \\
   \vspace{\tlim} \mbox{\scriptsize $t$ $\rightarrow$ $+$ $\infty$}
\end{array} \ e^{iH_{\ell} t} g(Q) e^{-iH t} F^{+}_{\ell}\p.
\end{eqnarray}

Let us show that the operator $\mathcal{W}$ is in fact the adjoint $(\Omega^{+}_{\ell})^*$ of the channel wave operator $\Omega^{+}_{\ell}$. Using the expressions (\ref{Identity Lg}) and (\ref{Separating LR3}) for $\Omega^{+}_{\ell}$ and $\mathcal{W}$ respectively one obtains the following useful identities by proceeding as in (\ref{F2=F}): $F^+_{\ell} \Omega^+_{\ell} = \Omega^+_{\ell}$ and $F^{+}_{0,\ell} \mathcal{W} = \mathcal{W}$. Then  
\begin{eqnarray*}
\langle (\Omega^{+}_{\ell})^* \psi | \varphi \rangle &=& \langle \psi | \Omega^{+}_{\ell} \varphi \rangle\\
&=& \langle F^{+}_{\ell} \psi | \Omega^{+}_{\ell} \varphi \rangle\\
&=& \lim_{t \rightarrow +\infty} \langle F^{+}_{0,\ell} e^{iH_{\ell} t} e^{-iHt} F^{+}_{\ell} \psi |  \varphi \rangle\\
&=& \langle F^{+}_{0,\ell} \mathcal{W} \psi |  \varphi \rangle\\
&=& \langle \mathcal{W} \psi |  \varphi \rangle\p.
\end{eqnarray*}
Hence $(\Omega^{+}_{\ell})^* = \mathcal{W}$.

We now collect the properties of the wave and scattering operators. 

(i) $\Omega^{\pm}_{a}$ is isometric on $\mathcal{H}^{\pm}_{0,a}$. Indeed, we have $\|\Omega^{\pm}_{a} \varphi\| = \lim_{t\rightarrow \pm \infty} \|e^{iHt} e^{-iH_{a}t} F^{\pm}_{0,a} \varphi\|$, which is equal to $\|\varphi\|$ if $\varphi \in \mathcal{H}^{\pm}_{0,a}$ and vanishes if $\varphi$ is orthogonal to $\mathcal{H}^{\pm}_{0,a}$. 

(ii) The range $\mathcal{R}(\Omega^{\pm}_{a})$ of $\Omega^{\pm}_{a}$ is the whole subspace $\mathcal{H}^{\pm}_{a}$. Indeed, let us consider for example the case of $\Omega^{+}_{\ell}$. Notice first that the identity $F^+_{\ell} \Omega^+_{\ell} = \Omega^+_{\ell}$ implies that $\mathcal{R}(\Omega^{+}_{\ell})$ is a subspace of $\mathcal{H}^{+}_{\ell}$. Then let $\psi \in \mathcal{H}^{+}_{\ell}$ and set $\varphi = \mathcal{W} \psi$. From the identity $F^{+}_{0,\ell} \mathcal{W} = \mathcal{W}$ one deduces that $\varphi$ belongs to $\mathcal{H}^{+}_{0,\ell}$. Finally we have $\psi = \Omega^{+}_{\ell} \varphi$ since
\begin{eqnarray*}
\| \Omega^{+}_{\ell} \varphi - \psi \| &=& \lim_{t \rightarrow +\infty} \| e^{iHt} e^{-iH_{\ell} t} F^{+}_{0,\ell} \varphi - \psi\|\\
&=& \lim_{t \rightarrow +\infty} \| \varphi -  e^{iH_{\ell}t} e^{-iH t} \psi\|\\
&=& \| \varphi -  \mathcal{W} \psi\| = 0\p.
\end{eqnarray*}

(iii) The wave operators $\Omega^{\pm}$ are isometries from $\mathcal{H}$ onto $\mathcal{H}_c(H)$ and therefore are complete, \ie $\mathcal{R}(\Omega^{\pm}) = \mathcal{H}_{c}(H)$, where $\mathcal{R}(\Omega^{\pm})$ denotes the range of $\Omega^{\pm}$. Indeed, writing $\Omega^{\pm} \equiv \Omega_{\ell}^{\pm} + \Omega_{r}^{\pm}$, using the points (i) and (ii), and recalling the decompositions  $\mathcal{H} = \mathcal{H}^{\pm}_{0,\ell} \oplus \mathcal{H}^{\pm}_{0,r}$ and $\mathcal{H}_{c}(H) = \mathcal{H}^{\pm}_{\ell} \oplus \mathcal{H}^{\pm}_{r}$ established in Section~\ref{Large time limits} it follows immediately that $\Omega^{\pm}$ is an isometry mapping $\mathcal{H}$ onto $\mathcal{H}_c(H)$.

(iv) The channel wave operators satisfy intertwining relations: $e^{-i H t} \Omega^{\pm}_{a} = \Omega^{\pm}_{a} e^{-i H_{a} t}$. More generally such relations hold for a large class of fonctions $\Phi$: $\Phi(H) \Omega^{\pm}_{a} = \Omega^{\pm}_{a} \Phi(H_{a})$. These relations express the conservation of energy (kinetic plus potential) in scattering processes (see the Remark in the next subsection).

(v) $S$ is unitary. Indeed, by (iii) we have 
\begin{eqnarray*}
S^{*} S &=& (\Omega^{-})^* \Omega^+ (\Omega^+)^* \Omega^-\\
&=& (\Omega^{-})^* F_c(H) \Omega^- = (\Omega^{-})^* \Omega^- = I
\end{eqnarray*}
and similarly $S S^{*} = I$, where $I$ denotes the identity operator in $\mathcal{H}$.

(vi) From (i) one easily sees that the operators $S_{a}$ and $S_{ab}$ are not isometries. However, it is clear that $S_{a}$ is isometric as an operator from $\mathcal{H}^{-}_{0,a}$ to $\mathcal{H}^{+}_{0,\ell} \oplus \mathcal{H}^{+}_{0,r} = \mathcal{H}$ and that the channel scattering operators $S_{ab}$ associate to each initial state in $\mathcal{H}^{-}_{0,b}$ a final state in $\mathcal{H}^{+}_{0,a}$. For example the operator $S_{\ell}$ maps initial states with positive momentum (corresponding to particles that are incident from the left) to the associated final states. If $\varphi$ is a wave packet with positive momentum, then $S_{\ell} \varphi$ is decomposed into the superposition of $S_{\ell \ell} \varphi$ (the part of the associated final state propagating towards the left) and $S_{r \ell} \varphi$ (the part propagating towards the right). The meaning of the operator $S_{r}$ is similar when applied to particles incident from the right.

(vii) The channel scattering operators satisfy the following important identities:
\begin{eqnarray}\label{Ortho}
(S_{ab})^* S_{cd} = 0, \hspace{5mm}  S_{ba} (S_{dc})^* = 0
\end{eqnarray}
if $a \not= c$, and
\begin{eqnarray}\label{Resolution Identity}
\sum_{a = \ell, r} (S_{ab})^* S_{ac} = \delta_{bc} F^-_{0,b}, \hspace{5mm} \sum_{a = \ell, r} S_{ba} (S_{ca})^* = \delta_{bc} F^+_{0,b}\p.
\end{eqnarray}
Indeed, (\ref{Ortho}) follow directly from the fact that $\Omega^+_{a} (\Omega^+_{c})^* = \delta_{ac} F^+_{a}$ and $\Omega^-_{a} (\Omega^-_{c})^* = \delta_{ac} F^-_{a}$. To check for example the first equation in (\ref{Resolution Identity}) we write 
\begin{eqnarray*}
\sum_{a = \ell, r} (S_{ab})^* S_{ac} &=&  (S_{\ell b})^* S_{\ell c} + (S_{rb})^* S_{rc}\\
&=& (\Omega_b^{-})^* \Omega_{\ell}^+ (\Omega_{\ell}^+)^* \Omega_c^- + (\Omega_b^{-})^* \Omega_{r}^+ (\Omega_{r}^+)^* \Omega_c^-\\
&=& (\Omega_b^{-})^*  [F^+_{\ell} + F^+_{r}] \Omega_c^-\\
&=& (\Omega_b^{-})^* F_c(H) \Omega_c^- = (\Omega_b^{-})^* \Omega_c^- = \delta_{bc} F^-_{0,b}\p.
\end{eqnarray*}

(viii) The intertwining relations for the channel wave operators immediately transcribe into the following intertwining relations for the channel scattering operators: $e^{-i H_{a} t} S_{ab} = S_{ab} e^{-i H_{b} t}$.

(ix) Let $\Theta$ be the time reversal operator given by complex conjugation: $(\Theta \psi)(x) = \overline{\psi(x)}$. It satisfies $\Theta F^{\pm}_{0,a} \Theta = F^{\mp}_{0,a}$, $\Theta F^{\pm}_{a} \Theta = F^{\mp}_{a}$ (see Section~\ref{Large time limits}) and $\Theta e^{iHt} e^{-iH_{a} t} F^{\pm}_{0,a} \Theta =  e^{-iHt} e^{iH_{a} t} F^{\mp}_{0,a}$, hence $\Theta \Omega^{\pm}_{a} \Theta = \Omega^{\mp}_{a}$. The last relation also implies that $\Theta (\Omega^{\pm}_{a})^* \Theta = (\Omega^{\mp}_{a})^*$. It follows that 
\begin{equation}\label{SAB}
\Theta S_{ab} \Theta = (S_{ba})^*\p,
\end{equation}
hence, recalling the decomposition $S = \sum_{a,b = \ell, r} S_{ab}$, one obtains  
\begin{equation}\label{S time reversal relation}
\Theta S \Theta = S^*\p.
\end{equation}

\subsection{Scattering matrix}\label{Scattering matrix}

As observed in Section~\ref{Large time limits} (see in particular (\ref{Projectors-ML}) and (\ref{Projectors-PR})) the subspace $\mathcal{H}_{+}$ of wave functions with positive momentum must be associated with different physical situations when one considers incoming or outgoing states. The same is true for the subspace $\mathcal{H}_{-}$ of wave functions with negative momentum. Wave functions in $\mathcal{H}_{+}$ (or $\mathcal{H}_{-}$) describe states that are incoming from the left (or the right), their evolution being governed by the Hamiltonian $H_{\ell}$ (or $H_{r}$). The component in $\mathcal{H}_{+}$ (or $\mathcal{H}_{-}$) of an outgoing state corresponds to a particle propagating towards the right (or the left), its evolution at large times being governed by $H_{r}$ (or $H_{\ell}$). Thus it is natural to introduce two free Hamiltonian $H^{\mathrm{in}}$ and $H^{\mathrm{out}}$ by
\begin{equation}\label{Free Hamiltonian}
H^{\mathrm{in}} = H_{\ell} \Pi_{+} + H_{r} \Pi_{-}, \hspace{5mm} H^{\mathrm{out}} = H_{r} \Pi_{+} + H_{\ell} \Pi_{-}
\end{equation}
($\Pi_{+}$ and $ \Pi_{-}$ being the orthogonal projections in $\mathcal{H}$ onto $\mathcal{H}_{+}$ and $\mathcal{H}_{-}$ respectively).

To diagonalize $H^{\mathrm{in}}$ we identify $\mathcal{H} = \mathcal{H}_{+} \oplus \mathcal{H}_{-}$ with a subspace $\mathcal{H}^{\mathrm{in}}$ of the complex Hilbert space $L^2((V_{\ell},\infty);\mathbb{C}^2)$ of square-integrable $2$-component functions of the variable $E \in (V_{\ell},\infty)$. The elements of $\mathcal{H}^{\mathrm{in}}$ have only one non-zero component for $E \in (V_{\ell},V_{r})$ and in general two non-zero components for $E > V_{r}$, their definition being as follows. Let  $\phi = \phi_{+} + \phi_{-}$ be a wave function ($\phi_{+}\in\mathcal{H}_{+}$, $\phi_{-}\in\mathcal{H}_{-}$). We denote by $\phi^{\mathrm{in}}(E) \in \mathbb{C}^2$ the value of $\phi$ at $E$ in $\mathcal{H}^{\mathrm{in}}$. In consideration of the meaning of $\phi_{\pm}$ when $\phi$ is viewed as an incoming wave packet, we use the notations $\phi^{\mathrm{in}}_{\ell}(E)$ for the component of $\phi^{\mathrm{in}}(E)$ associated to $\phi_{+}$ and $\phi^{\mathrm{in}}_{r}(E)$ for that associated to $\phi_{-}$. So 
\begin{equation}\label{Wave In}
\phi^{\mathrm{in}}(E)=\begin{pmatrix} \phi^{\mathrm{in}}_{\ell}(E) \\  \phi^{\mathrm{in}}_{r}(E)
\end{pmatrix}
\end{equation}
with
\begin{eqnarray}
\phi^{\mathrm{in}}_{\ell}(E) &=&  \frac{1}{[4(E-V_{\ell})]^{1/4}} \widehat{\phi_{+}}(\sqrt{E-V_{\ell}}) \hspace{5mm} (E > V_{\ell})\label{State-L}\\
\phi^{\mathrm{in}}_{r}(E) &=&  \frac{1}{[4(E-V_{r})]^{1/4}} \widehat{\phi_{-}}(-\sqrt{E-V_{r}}) \hspace{5mm} (E > V_{r})\label{State-R}
\end{eqnarray}
and $\phi^{\mathrm{in}}_{r}(E) = 0$ for $E \leq V_{r}$. The normalization factors on the \rhs of (\ref{State-L}) and  (\ref{State-R}) are chosen such that the identification of $\mathcal{H}$ with the subspace $\mathcal{H}^{\mathrm{in}}$ of $L^2((V_{\ell},\infty);\mathbb{C}^2)$ is unitary. Clearly $(H^{\mathrm{in}}\phi)^{\mathrm{in}}(E) = E \phi^{\mathrm{in}}(E)$.

The diagonalization of $H^{\mathrm{out}}$ is achieved similarly by identifying $\mathcal{H}$ with a subspace $\mathcal{H}^{\mathrm{out}}$ of $L^2((V_{\ell},\infty);\mathbb{C}^2)$ in the following manner: for $\phi = \phi_{+} + \phi_{-}$ interpreted as an outgoing state ($\phi_{+}\in\mathcal{H}_{+}$, $\phi_{-}\in\mathcal{H}_{-}$) we set
\begin{eqnarray}
\phi^{\mathrm{out}}_{r}(E) &=&  \frac{1}{[4(E-V_{r})]^{1/4}} \widehat{\phi_{+}}(\sqrt{E-V_{r}}) \hspace{5mm} (E > V_{r})\label{State-R Out}\\
\phi^{\mathrm{out}}_{\ell}(E) &=&  \frac{1}{[4(E-V_{\ell})]^{1/4}} \widehat{\phi_{-}}(-\sqrt{E-V_{\ell}}) \hspace{5mm} (E > V_{\ell})\label{State-L Out}
\end{eqnarray}
and 
\begin{equation}\label{Wave Out}
\phi^{\mathrm{out}}(E)=\begin{pmatrix} \phi^{\mathrm{out}}_{r}(E) \\  \phi^{\mathrm{out}}_{\ell}(E) 
\end{pmatrix}\p,
\end{equation}
with $\phi^{\mathrm{out}}_{r}(E) = 0$ if $E \leq V_{r}$.

From the relations $S_{ab} H_{b} = H_{a} S_{ab}$ one finds that
\begin{equation}\label{S H}
S H^{\mathrm{in}} = H^{\mathrm{out}} S\p.
\end{equation}
If $S$ is viewed as an operator from $\mathcal{H}^{\mathrm{in}}$ to $\mathcal{H}^{\mathrm{out}}$, (\ref{S H}) means that $S$ maps the value at energy $E$ of an incoming state to the value of the associated outgoing state at the \emph{same} energy $E$: for each $E > V_{r}$ there is a (unitary) $2\,\mbox{x}\,2$ matrix $S(E)$, called the \emph{scattering matrix}, such that
\begin{equation}\label{Out In}
(S\varphi)^{\mathrm{out}}(E) = S(E) \varphi^{\mathrm{in}}(E)
\end{equation}
or more explicitly
\begin{equation}\label{Smatrix}
\begin{pmatrix} (S\varphi)^{\mathrm{out}}_{r}(E) \\  (S\varphi)^{\mathrm{out}}_{\ell}(E) 
\end{pmatrix} =  \begin{pmatrix}
  S_{r \ell}(E) & S_{rr}(E) \\
  S_{\ell \ell}(E) & S_{\ell r}(E)
\end{pmatrix} \begin{pmatrix} \varphi^{\mathrm{in}}_{\ell}(E) \\  \varphi^{\mathrm{in}}_{r}(E)
\end{pmatrix}\p.
\end{equation}
For $V_{\ell} < E < V_{r}$, where $\varphi^{\mathrm{in}}_{r}(E) = (S\varphi)^{\mathrm{out}}_{r}(E) = 0$, there is a complex number $S_{\ell \ell}(E)$ of modulus 1 such that $(S\varphi)^{\mathrm{out}}_{\ell}(E)=S_{\ell \ell}(E) \varphi^{\mathrm{in}}_{\ell}(E)$. The relation (\ref{Out In}) specifies the wave function $S\varphi$, viewed as an element of $\mathcal{H}^{\mathrm{out}}$, in terms of the wave function $\varphi$ represented in $\mathcal{H}^{\mathrm{in}}$.

We point out a simple way of arriving at the above structure. Consider the operator $\mathcal{J}$ in $\mathcal{H}$ that interchanges $\mathcal{H}_{+}$ and $\mathcal{H}_{-}$ (\ie the parity operator):
\begin{equation}
(\widehat{\mathcal{J}\phi})(p) = \hat{\phi}(-p)\p.
\end{equation}
It satisfies $\mathcal{J}^{*}=\mathcal{J}$, $\mathcal{J}^2=I$, $\mathcal{J} \mathcal{H}_{\pm}=\mathcal{H}_{\mp}$, and it interwines $H^{\mathrm{in}}$ and $H^{\mathrm{out}}$:
\begin{equation}
H^{\mathrm{in}} \mathcal{J} = \mathcal{J} H^{\mathrm{out}}\p.
\end{equation}
It follows that $H^{\mathrm{in}}\mathcal{J}S=\mathcal{J}H^{\mathrm{out}}S=\mathcal{J}SH^{\mathrm{in}}$, \ie $\mathcal{J}S$ commutes with $H^{\mathrm{in}}$. So (see \eg Proposition 5.27 in \cite{AJS}), in the representation $\mathcal{H}^{\mathrm{in}}$ diagonalizing the self-adjoint operator $H^{\mathrm{in}}$, the unitary operator $\mathcal{J}S$ is decomposable, \ie for example for each $E > V_{r}$ there is a unitary $2\,\mbox{x}\,2$ matrix $\sigma(E)$ such that for each $\varphi$ in $\mathcal{H}$:
\begin{equation}
(\mathcal{J} S \varphi)^{\mathrm{in}}(E) = \sigma(E) \varphi^{\mathrm{in}}(E)\p.
\end{equation}
It is easy to check that, if written as an operator $\mathcal{J}_{\mathrm{in} \rightarrow \mathrm{out}}$ from $\mathcal{H}^{\mathrm{in}}$ to $\mathcal{H}^{\mathrm{out}}$, $\mathcal{J}$ is given by
\begin{equation}\label{Eq J}
(\mathcal{J}\phi)^{\mathrm{out}}(E) \equiv (\mathcal{J}_{\mathrm{in} \rightarrow \mathrm{out}} \phi)^{\mathrm{out}}(E) =  \begin{pmatrix}
  0 & 1 \\
  1 & 0
\end{pmatrix} \phi^{\mathrm{in}}(E)\p.
\end{equation}
Hence, writing the scattering operator $S : \mathcal{H}^{\mathrm{in}} \rightarrow \mathcal{H}^{\mathrm{out}}$ as $S = \mathcal{J}(\mathcal{J} S) = \mathcal{J}_{\mathrm{in} \rightarrow \mathrm{out}} (\mathcal{J}S)_{\mathrm{in} \rightarrow \mathrm{in}}$, one sees that  (for $E > V_{r}$) $S$ has the form (\ref{Smatrix}), with $S(E) =  \begin{pmatrix}
  0 & 1 \\
  1 & 0
\end{pmatrix} \sigma(E)$ a unitary $2\,\mbox{x}\,2$ matrix.

We add a few comments concerning the $S$-matrix $S(E)$ in (\ref{Smatrix}). For $E > V_r$ the complex numbers $S_{\ell \ell}(E)$ and  $S_{r r}(E)$ are the reflection amplitudes at energy $E$, whereas $S_{r \ell}(E)$ and  $S_{\ell r}(E)$ represent the transmission amplitudes. Time reversal symmetry and unitarity lead to important relations between these quantities. Equation (\ref{S time reversal relation}) may be rewritten as $S = \Theta S^{*} \Theta$. As an operator from $\mathcal{H}^{\mathrm{out}}$ to $\mathcal{H}^{\mathrm{in}}$, $S^{*}$ is decomposable, its component at energy $E > V_r$ is just
\begin{equation}\label{Smatrix adjoint}
S^{*}(E) = S(E)^{*} = \begin{pmatrix}
  \overline{S_{r \ell}(E)} & \overline{S_{\ell \ell}(E)} \\
  \overline{S_{rr}(E)} & \overline{S_{\ell r}(E)}
\end{pmatrix}\p.
\end{equation}
The time reversal operator $\Theta$, expressed as an operator from $\mathcal{H}^{\mathrm{in}}$ to $\mathcal{H}^{\mathrm{out}}$, is also decomposable ($\mathcal{C}$ denotes complex comjugation):

\begin{equation*}
(\Theta \phi)^{\mathrm{out}}(E) =  \begin{pmatrix}
  0 & \mathcal{C} \\
  \mathcal{C} & 0
\end{pmatrix} \phi^{\mathrm{in}}(E) = \begin{pmatrix} \overline{\phi^{\mathrm{in}}_{r}(E)} \\  \overline{\phi^{\mathrm{in}}_{\ell}(E)} 
\end{pmatrix}\p.
\end{equation*}
So, writing the operator $\Theta S^{*} \Theta : \mathcal{H}^{\mathrm{in}} \rightarrow \mathcal{H}^{\mathrm{out}}$ as  $\Theta_{\mathrm{in} \rightarrow \mathrm{out}} (S^{*})_{\mathrm{out} \rightarrow \mathrm{in}} \Theta_{\mathrm{in} \rightarrow \mathrm{out}}$, the equation $S=\Theta S^{*} \Theta$ implies that
\begin{equation*}
\begin{pmatrix}
  S_{r \ell}(E) & S_{rr}(E) \\
  S_{\ell \ell}(E) & S_{\ell r}(E)
\end{pmatrix} = \begin{pmatrix}
  S_{\ell r}(E) & S_{rr}(E) \\
  S_{\ell \ell}(E) & S_{r \ell}(E)
\end{pmatrix}\p,
\end{equation*}
hence
\begin{equation}
S_{r \ell}(E) = S_{\ell r}(E)\p.
\end{equation}
The transmission amplitudes from left to right and from right to left are equal. The unitarity of $S(E)$ then leads to 
\begin{equation}
|S_{\ell \ell}(E)|^2 = 1 - |S_{r \ell}(E)|^2 = |S_{r r}(E)|^2\p,
\end{equation}
\ie the reflection probabilities at energy $E > V_r$ from the left and from the right are the same.

 \textbf{Remark.} The variable $E$ on the \rhs of (\ref{Out In}) refers to the value of $H^{\mathrm{in}}$ whereas that on the \lhs relates to $H^{\mathrm{out}}$. As an illustration consider a scattering state $\psi$ in $\mathcal{H}^{-}_{\ell}$ (\ie incoming from the left) having energy support (with respect to $H$) in a very small interval $\Delta$ centered at $E_0 > V_r$. We shall say that $\psi$ is a state at energy $E_0$. By the intertwining relation  $(\Omega^{-}_{\ell})^{*} \Phi(H) = \Phi(H_{\ell}) (\Omega^{-}_{\ell})^{*}$ the associated initial state $\varphi=(\Omega^{-}_{\ell})^{*} \psi$ in  $\mathcal{H}^{-}_{0,\ell}$ is a state at free energy $E_0$ (relative to $H^{\mathrm{in}}$, \ie $\varphi^{\mathrm{in}}(E) \equiv \varphi^{\mathrm{in}}_{\ell}(E) = 0$ for $E$ outside $\Delta$); for this state it is natural to view $E_0$ as being composed of a potential energy $V_\ell$ and a kinetic energy $\lambda_\ell \approx E_{0} - V_\ell$. Likewise the components  $S_{\ell \ell}\varphi \equiv \Omega^{+}_{\ell} \psi$ and $S_{r \ell}\varphi \equiv \Omega^{+}_{r} \psi$ of the final state $S_{\ell}\varphi$ are at free energy $E_0$ (relative to $H^{\mathrm{out}}$), \ie $(S_{\ell \ell}\varphi)^{\mathrm{out}}(E) = 0$ and $(S_{r \ell}\varphi)^{\mathrm{out}}(E) = 0$ for $E \not \in \Delta$, their kinetic energy being $\lambda_\ell \approx E_{0} - V_\ell$  and $\lambda_r \approx E_{0} - V_r$ respectively. This expresses the fact that reflection is an elastic process while transmission is inelastic if $V_\ell \not = V_r$ (the operator $S_{\ell \ell}$ commutes with $H_0 = P^2$ whereas $S_{r \ell}$ does not if $V_\ell \not = V_r$).

\section{Time delay}\label{Time delay}

As explained in the Introduction time delay expresses the excess time that
scattered particles spend in the scattering region when compared to free
particles and therefore is naturally
formalized, in the framework of time-dependent scattering theory, in terms of
sojourn times. However one often uses in calculations the Eisenbud-Wigner
expression of time delay given in terms of the $S$-matrix. In this section we prove that these two representations of
time delay are in fact identical. We first show that for appropriate initial
states $\varphi$ the sojourn times
(\ref{Sojourn-Time})-(\ref{Sojourn-Time-Out}) are finite for each finite $R$,
so that the local time delays (\ref{Time-Delay-In})-(\ref{Time-Delay-Average})
are well defined. We then show that in the case $V_{\ell} \not= V_{r}$ only
the symmetrized expression (\ref{Time-Delay-Average}) of local time delay has
a finite limit as $R \rightarrow \infty$; we call this limit the \emph{global time delay}. In the context of time-independent scattering theory we first observe that the Eisenbud-Wigner operator $\mathcal{T} = \{\mathcal{T}(E)\}$ is well defined and self-adjoint, so that the Eisenbud-Wigner expression of time delay (\ref{EW Time Delay}) is quantum-mechanically natural. We finally establish the main result of this paper, namely the identity between the global time delay and the Eisenbud-Wigner expression of time delay.

We assume that $\mu > 2$ in the decay assumptions (\ref{Assumption VL})-(\ref{Assumption VR}) on the potential. In order to avoid longer expressions, we discuss the case of wave packets that are incident from the left ($\varphi \in \mathcal{H}^{-}_{0,\ell} = \mathcal{H}_{+}$). The same type of arguments are applicable to wave packets incident from the right ($\varphi \in \mathcal{H}^{-}_{0,r} = \mathcal{H}_{-}$) and to general initial states ($\varphi \in \mathcal{H}$). More specifically we shall consider initial states $\varphi \in \mathcal{H}_{+}$ having energy support (with respect to $H_{\ell}$) away from the thresholds $V_\ell$ and $V_r$, and some decay in configuration space. For this we introduce a parameter $\theta \geq 0$ (its values will be specified further on) and denote by $\mathcal{D}^{\mathrm{in}}_{\theta}$ the set of wave functions $\varphi \in \mathcal{H}_{+}$ satisfying
\begin{itemize}
\item[(i)] there are intervals $\Delta_1 = [E_1,E'_1]$ and  $\Delta_2 = [E_2,E'_2]$ (depending on $\varphi$) with $V_{\ell}<E_1<E'_1<V_r<E_2<E'_2<\infty$ such that $\varphi$ has energy support (with respect to $H_{\ell}$) in $\Delta_1 \cup \Delta_2$,
\item[(ii)] the following integrability condition:
\begin{equation}\label{Decay Wave Packet}
\int_{-\infty}^{\infty} |(1+|x|)^{\theta} \varphi(x)|^2 \, dx < \infty\p.
\end{equation}
\end{itemize}

To handle the final states $S_{\ell}\varphi$ (which evolve with $H^{\mathrm{out}}$) we introduce a similar set of wave functions $\mathcal{D}^{\mathrm{out}}_{\theta}$ as follows: $\phi = \phi_+ + \phi_- \in \mathcal{H}_{+} \oplus \mathcal{H}_{-}$ belongs to $\mathcal{D}^{\mathrm{out}}_{\theta}$ if there are intervals $\Delta_1$ and $\Delta_2$ as in (i) above such that: (1) $\phi_+$ has energy support with respect to $H_{r}$ in $\Delta_2$ and $\varphi_-$ has energy support with respect to $H_{\ell}$ contained in $\Delta_1 \cup \Delta_2$, (2) $\phi_+$ and $\phi_-$ satisfy the integrability condition (\ref{Decay Wave Packet}). Notice that, if $\varphi \in \mathcal{H}_+$ has the support property (i) in the definition of $\mathcal{D}^{\mathrm{in}}_{\theta}$, then $\phi \equiv S_\ell \varphi$ satisfies the support condition in $\mathcal{D}^{\mathrm{out}}_{\theta}$ (this follows from the intertwining relations $H_{a} S_{ab} = S_{ab}H_{b}$, observing that $\phi_+ = S_{r\ell}\varphi$ and $\phi_- = S_{\ell\ell}\varphi$). Nevertheless some assumptions on the decay of $V$ are needed so that $S_\ell$ maps $\mathcal{D}^{\mathrm{in}}_{\theta}$ into $\mathcal{D}^{\mathrm{out}}_{\theta}$ (see Section~\ref{Local time delay}).

\subsection{Sojourn times}\label{Sojourn time}

For $0 < R < \infty$ the sojourn times in the interval $[-R,R]$ associated to a wave function $\varphi \in \mathcal{H}_{+}$ are defined by (see (\ref{Sojourn-Time})-(\ref{Sojourn-Time-Out}))
\begin{eqnarray}
T_{R}(\Omega^{-} \varphi) &=& \int_{-\infty}^{\infty} dt \int_{-R}^{R} dx \ |(e^{-iHt} \Omega^{-}_{\ell} \varphi)(x)|^2\p.\label{Sojourn-Time2}\\
T^{\mathrm{in}}_{R}(\varphi) &=& \int_{-\infty}^{\infty} dt \int_{-R}^{R} dx \ |(e^{-iH_{\ell}t} \varphi)(x)|^2\p,\label{Sojourn-Time-In2}\\
T^{\mathrm{out}}_{R}(\varphi)&=& \int_{-\infty}^{\infty} dt \int_{-R}^{R} dx \ |(e^{-iH_{\ell}t} S_{\ell \ell} \varphi)(x) + (e^{-iH_{r}t} S_{r \ell} \varphi)(x)|^2\p,\label{Sojourn-Time-Out2}
\end{eqnarray}
If in addition $\varphi$ has energy support (with respect to $H_{\ell}$) away from $V_{\ell}$ and $V_r$ these quantities are finite. This is not surprising since such wave functions describe states of a particle with non-zero velocity. Mathematically the finiteness of $T_{R}( \Omega^{-} \varphi)$ follows from (\ref{Propagation-Estimate-Strong-Consequence}) with $f(x)=1$ or $0$ if $|x| \leq R$ or $|x| > R$ respectively ($\psi = \Omega^{-} \varphi$ has energy support with respect to $H$ away from $V_{\ell}$ and $V_r$ by the intertwining relation $H \Omega^{-}_{\ell} = \Omega^{-}_{\ell} H_{\ell}$, and $|f(x)| \leq (R+1) (1+|x|)^{-1}$). The finiteness of $T^{\mathrm{in}}_{R}(\varphi)$ and  $T^{\mathrm{out}}_{R}(\varphi)$ can be obtained in the same way, applying (\ref{Propagation-Estimate-Strong-Consequence}) for the Hamiltonians $H_{\ell}$ and $H_r$, with $\psi_t = e^{-iH_{\ell}t}\varphi$, $\psi_t = e^{-iH_{\ell}t} S_{\ell \ell}\varphi$ and $\psi_t = e^{-iH_{r}t} S_{r \ell}\varphi$. 

If $\varphi \in \mathcal{H}_+$ is a wave packet satisfying (\ref{Decay Wave Packet}) for some $\theta > 1$, the finiteness of $T^{\mathrm{in}}_{R}(\varphi)$ is also an immediate  consequence of the decay estimates (\ref{Propagation-a}) and (\ref{Propagation-a2}) (take $x_0 = R$ in (\ref{Propagation-a}) and $x_0 = -R$ in  (\ref{Propagation-a2})). Similarly the finiteness of $T^{\mathrm{out}}_{R}(\varphi)$ then follows from (\ref{Propagation-a})-(\ref{Propagation-b2}) provided that one knows that $S_{\ell \ell} \varphi$ and  $S_{r \ell} \varphi$ also have the decay property (\ref{Decay Wave Packet}) with $\theta > 1$. 

\subsection{Local time delay}\label{Local time delay}

The results of the preceding subsection imply that, for initial states belonging to $\mathcal{D}^{\mathrm{in}}_{\theta}$ with $\theta \geq 0$, the local time delays (\ref{Time-Delay-In})-(\ref{Time-Delay-Average}) are finite for each finite $R$. We now turn to the question of existence of a limit of these quantities as $R \rightarrow \infty$. Following \cite{AC} we proceed in two steps: (1) approximate $\tau^{\mathrm{in}}_{R}(\varphi)$ and  $\tau^{\mathrm{out}}_{R}(\varphi)$ by expressions giving the same limit but involving the scattering operator $S_{\ell}$ rather than the wave operator $\Omega^{-}_{\ell}$, (2) use an asymptotic expansion (for large $R$) of $\int_{0}^{\pm \infty} e^{iH_{\kappa}t} \chi_{(-R,R)}(Q) e^{-iH_{\kappa}t} dt$, where $ \chi_{(-R,R)}(x)=1$ or $0$ if $|x| < R$ or  $|x| \geq R$ respectively. We treat here step (1) and discuss step (2) in the next subsection. We assume that $\varphi$ belongs to $\mathcal{D}^{\mathrm{in}}_{\theta}$ for some $\theta > 4$.

Let us consider $\tau^{\mathrm{in}}_{R}(\varphi)$. Setting $\varphi_t = e^{-iH_{\ell}t} \varphi$ we have 
\begin{equation}\label{Tin}
\tau^{\mathrm{in}}_{R}(\varphi) = \int_{-\infty}^{\infty} dt \int_{-R}^{R} dx \left[
  |(\Omega^{-}_{\ell} \varphi_t)(x)|^2 - |\varphi_t(x)|^2\right] \equiv \int_{-\infty}^{\infty} I_{R}(t) \, dt\p,
\end{equation}
where $I_{R}(t)$ denotes the integral over the variable $x$. Observe that for any finite $t_0$ and $t_1$ ($t_0 < t_1$), the quantities $\int_{t_0}^{t_1} dt \int_{-R}^{R} dx \, |(\Omega^{-}_{\ell} \varphi_t)(x)|^2$ and  $\int_{t_0}^{t_1} dt \int_{-R}^{R} dx \, |\varphi_t(x)|^2$ are increasing functions of $R$ each of which converges to $(t_1-t_0) \int_{-\infty}^{\infty} |\varphi(x)|^2 \, dx$ as $R \rightarrow \infty$.  Hence, for any $t_0 < t_1$:
\begin{equation}\label{Tin 3 parts}
\tau^{\mathrm{in}}_{R}(\varphi) = \int_{-\infty}^{t_0} I_{R}(t) \, dt + \int_{t_1}^{\infty} I_{R}(t) \, dt + {\scriptstyle \cal O}(1) \mbox{ as } R \rightarrow \infty\p.
\end{equation}
Now, by the definition of the wave operators, $\Omega^{-}_{\ell} \varphi_t$ and $\varphi_t$ approach each other (in the Hilbert space norm) as $t \rightarrow -\infty$.  One thus expects that the integral over $(-\infty,t_0)$ in (\ref{Tin 3 parts}) is negligible if $t_0$ is negative and sufficiently large, so that it suffices to consider the contribution to $\tau^{\mathrm{in}}_{R}(\varphi)$ coming from large positive times. As explained below this is indeed the case.

 Denoting by $\chi_{(-R,R)}$ multiplication by $\chi_{(-R,R)}(x)$, writing $I_R(t)$ in terms of inner products and then using the Cauchy-Schwarz inequality we have
\begin{eqnarray}\label{EqC}
|I_R(t)| &=& |\langle \Omega^{-}_{\ell} \varphi_t | \chi_{(-R,R)} (\Omega^{-}_{\ell} \varphi_t - \varphi_t)\rangle + \langle \Omega^{-}_{\ell} \varphi_t - \varphi_t | \chi_{(-R,R)} \varphi_t \rangle|\nonumber\\
&\leq& \|\Omega^{-}_{\ell} \varphi_t\| \cdot \| \chi_{(-R,R)} (\Omega^{-}_{\ell} \varphi_t - \varphi_t)\| + \|\Omega^{-}_{\ell} \varphi_t - \varphi_t\| \cdot \|\chi_{(-R,R)} \varphi_t\|\nonumber\\
&\leq& 2 \|\varphi\| \cdot \|(\Omega^{-}_{\ell}-I) \varphi_t\|\p.
\end{eqnarray}
We now observe that
\begin{eqnarray}
(\Omega^{-}_{\ell} - I) \varphi_t &=&  - \int_{-\infty}^{0} \frac{d}{du}\left[e^{iHu} e^{-iH_{\ell}u}\right] e^{-iH_{\ell}t} \varphi \, du\nonumber\\
&=& - i e^{-iHt} \int_{-\infty}^{t} e^{iHs} (V-V_{\ell}) e^{-iH_{\ell}s}
\varphi \, ds\p.
\end{eqnarray}
By taking into account the (continuous) triangle inequality in the Hilbert space norm, one obtains that
\begin{eqnarray}\label{EqE}
\|(\Omega^{-}_{\ell} - I) \varphi_t \| &\leq& \int_{-\infty}^{t} \|(V-V_{\ell})\varphi_s\| \, ds\nonumber\\
&\leq&  \int_{-\infty}^{t} \|(V-V_{\ell})\chi_{\ell}\varphi_s\| \, ds + \int_{-\infty}^{t} \|(V-V_{\ell})\chi_{r}\varphi_s\| \, ds\p.
\end{eqnarray}
The norms appearing in the integrands can be estimated (for $s < 0$) by using (\ref{Propagation-c}) and (\ref{Propagation-a2}):
\begin{eqnarray*}
\|(V-V_{\ell})\chi_{\ell}\varphi_s\| &\leq& C (1+|s|)^{-\rho}\\
\|(V-V_{\ell})\chi_{r}\varphi_s\| &\leq& C (1+|s|)^{-\theta/2}\p.
\end{eqnarray*}
Here $C$ is some constant (depending on $\varphi$ and $\theta$) and $\rho=\min\{\mu,\theta/2\}$. It then follows from (\ref{EqC}) and  (\ref{EqE}) that, for some constant $\tilde{C}$ and $t < 0$:
\begin{equation}\label{t negative H H0}
|I_R(t)| \leq \tilde{C} |t|^{1-\rho}\p.
\end{equation}
This bound is valid for all $R > 0$ (the constant $\tilde{C}$ depends on $\varphi$ and $\theta$ but is independent of $R$). Since we assumed that $\theta > 4$ and $\mu > 2$, we have $\rho > 2$, so that $|\int_{-\infty}^{t_0} I_R(t) dt|$ can be made arbitrarily small, independently of $R$, by choosing $t_0$ sufficiently large (negative).

As $t \rightarrow +\infty$, $\Omega^{-}_{\ell} \varphi_t$ approaches $S_{\ell} \varphi_t$. Writing $\Omega^{-}_{\ell} \varphi_t - S_{\ell} \varphi_t = (\Omega^{+}_{\ell} - I) S_{\ell \ell} \varphi_t + (\Omega^{+}_{r} - I) S_{r \ell} \varphi_t$ and proceeding as above (using the decay estimates  (\ref{Propagation-a}), (\ref{Propagation-b}) and (\ref{Propagation-c})) one finds that
\begin{equation}\label{Arbitrarily small}
\left|\int_{t_1}^{\infty} dt \int_{-R}^{R} dx \left[|(\Omega^{-}_{\ell} \varphi_t)(x)|^2 - |(S_{\ell} \varphi_t)(x)|^2 \right] \right|
\end{equation}
can be made arbitrarily small, independently of $R$, by choosing $t_1 > 0$ large enough, under the proviso that $S_{\ell}\varphi$ belongs to $\mathcal{D}^{\mathrm{out}}_{\theta}$ for some $\theta > 4$. Hence, setting 
\begin{equation}\label{Sigma In}
\sigma^{\mathrm{in}}_{R}(\varphi) = \int_{0}^{\infty} dt \int_{-R}^{R} dx \left[
  |(S_{\ell}\varphi_t)(x)|^2 - |\varphi_t(x)|^2\right]
\end{equation}
we have
\begin{equation}\label{Tau Sigma In}
\lim_{R \rightarrow \infty} \tau^{\mathrm{in}}_{R}(\varphi) = \lim_{R \rightarrow \infty} \sigma^{\mathrm{in}}_{R}(\varphi)\p.
\end{equation}
The relation (\ref{Tau Sigma In}) shows that, as $R \rightarrow \infty$, either $\tau^{\mathrm{in}}_{R}(\varphi)$ and $\sigma^{\mathrm{in}}_{R}(\varphi)$ converge to the same finite limit or both are diverging.

By the same type of arguments one finds that 
\begin{equation}\label{Tau Sigma Out}
\lim_{R \rightarrow \infty} \tau^{\mathrm{out}}_{R}(\varphi) = \lim_{R \rightarrow \infty} \sigma^{\mathrm{out}}_{R}(\varphi)\p,
\end{equation}
with
\begin{equation}\label{Sigma Out}
\sigma^{\mathrm{out}}_{R}(\varphi) = -\int_{-\infty}^{0} dt \int_{-R}^{R} dx \left[
  |(S_{\ell}\varphi_t)(x)|^2 - |\varphi_t(x)|^2\right]\p.
\end{equation}

The relations (\ref{Tau Sigma In}) and (\ref{Tau Sigma Out}) are satisfied for wave functions $\varphi$ belonging to $\mathcal{D}^{\mathrm{in}}_{\theta}$ for some $\theta > 4$ such that $S_{\ell} \varphi$ belongs to $\mathcal{D}^{\mathrm{out}}_{\theta}$ (for some possibly different value of $\theta > 4$). This restriction essentially amounts to a differentiability condition on the $S$-matrix $S(E)$, namely that $S(E)$ should be $\theta$ times differentiable (with respect to $E$) away from the thresholds $V_{\ell}$ and $V_{r}$. Differentiability of $S(E)$ can be obtained for example from differentiability assumptions on the potential $V$ or from assumptions on the decay of $V$ at large $|x|$ (\ie assumptions on the parameter $\mu$ occuring in (\ref{Assumption VL})-(\ref{Assumption VR})). The entries of the matrix $S(E)$ are simple expressions in terms of the Jost solutions of the stationary Schr\"{o}dinger equation, and it suffices to know that the Jost solutions are $\theta$ times differentiable with respect to the energy parameter. We refer to Appendix~D for a brief discussion of these questions. The simplest situation is obtained by taking $\theta = 5$. Then the above-mentioned conditions on $\varphi$ and $S_{\ell}\varphi$ are satisfied if $\varphi$ fulfills the support condition (i) in the definition of $\mathcal{D}^{\mathrm{in}}_{\theta}$ and $\hat{\varphi}$ is five times continuously differentiable, and if $\mu > 6$ in  (\ref{Assumption VL})-(\ref{Assumption VR}) (then $S(E)$ is five times continuously differentiable away from the thresholds). If one admits differentiability of fractional order (Lipschitz or H\"{o}lder continuity of $\hat{\varphi}$ and $\widehat{S\varphi}$), then it suffices to assume that $\mu > 5$. If $\mu > 5$, then the fourth derivative $S^{(4)}(E)$ of $S(E)$ is locally H\"{o}lder continuous with exponent $\gamma = \min\{1,\mu-5\}$, \ie $|S^{(4)}_{ab}(E) - S^{(4)}_{ab}(E')| \leq C_{\Delta} |E-E'|^{\gamma}$ for $E,E'$ in any closed energy interval $\Delta$ not containing the thresholds. If $\hat{\varphi}^{(4)}$ is also H\"{o}lder continuous with the same exponent $\gamma$, then for $\delta < \gamma$:
\begin{equation}\label{Sphi Dtheta}
\int_{-\infty}^{\infty} |(1+|x|)^{\delta} (1+|x|)^4 (S_{ab} \varphi)(x)|^2 \, dx < \infty\p,
\end{equation}
\ie $S_{\ell} \varphi \in \mathcal{D}^{\mathrm{out}}_{4+\delta}$. The finiteness of the integral in (\ref{Sphi Dtheta}) follows from a result in classical Fourier analysis stating that, if $f \in L^2(\mathbb{R})$, $\hat{f}(p)=0$ for all $p$ outside some bounded subset of $\mathbb{R}$ and $|\hat{f}(p) - \hat{f}(p')| \leq C |p-p'|^{\gamma}$ for some $0 < \gamma \leq 1$, then $\int_{-\infty}^{\infty} |(1+|x|)^{\delta} f(x)|^2 \, dx < \infty$ for $\delta < \gamma$ (see Section~4.13 in \cite{Ti}, in particular Theorem 85 and its proof).

\subsection{Global and Eisenbud-Wigner time delay}\label{Global and Eisenbud-Wigner time delay}

Thoughout this subsection we assume that, for some $\theta > 4$, $\varphi$ is a wave packet in $\mathcal{D}^{\mathrm{in}}_{\theta}$ such that  $S_{\ell} \varphi$ belongs to $\mathcal{D}^{\mathrm{out}}_{\theta}$. Then, as seen in the preceding subsection, one can study the limit of the local time delays $\tau^{\mathrm{in}}_{R}(\varphi)$ and $\tau^{\mathrm{out}}_{R}(\varphi)$ as $R \rightarrow \infty$ by considering the limit of the quantities $\sigma^{\mathrm{in}}_{R}(\varphi)$ and $\sigma^{\mathrm{out}}_{R}(\varphi)$, see (\ref{Sigma In})-(\ref{Sigma Out}). Using the identity $(S_{\ell})^{*} S_{\ell} = \Pi_{+}$ one can rewrite $\sigma^{\mathrm{in}}_{R}(\varphi)$ as follows ($\varphi \in \mathcal{H}_{+}$, $\varphi_t = e^{-iH_\ell t}\varphi$): 
\begin{equation}
\sigma^{\mathrm{in}}_{R}(\varphi) = \int_{0}^{\infty} \langle S_{\ell} \varphi_t |
[\chi_{(-R,R)},S_{\ell}] \varphi_t \rangle \, dt\p.
\end{equation}
Next, writing explicitly $S_{\ell}=S_{\ell \ell} + S_{r \ell}$ and using the orthogonality relations (\ref{Ortho}) as well as the 
intertwining relations $e^{-iH_{a}t} S_{ab} =  S_{ab} e^{-iH_{b}t}$ one obtains
\begin{eqnarray}\label{Integral 4}
\sigma^{\mathrm{in}}_{R}(\varphi) &=& \int_{0}^{\infty} \langle S_{\ell \ell} \varphi | [e^{iH_0 t} \chi_{(-R,R)} e^{-iH_0 t}, S_{\ell \ell}] \varphi \rangle \, dt\nonumber\\
 &+& \int_{0}^{\infty} \langle  S_{r \ell} \varphi | [e^{iH_0 t} \chi_{(-R,R)}
 e^{-iH_0 t}, S_{r \ell}] \varphi \rangle \, dt\nonumber\\
&+& \int_{0}^{\infty} \langle e^{-iH_{\ell}t} S_{\ell \ell} \varphi |  \chi_{(-R,R)}
 e^{-iH_{r}t} S_{r \ell} \varphi \rangle \, dt\nonumber\\
&+& \int_{0}^{\infty} \langle e^{-iH_{r}t} S_{r \ell} \varphi |  \chi_{(-R,R)}
 e^{-iH_{\ell}t} S_{\ell \ell} \varphi \rangle \, dt\p.
\end{eqnarray}
In the limit $R \rightarrow \infty$ the last two integrals in (\ref{Integral 4}) vanish. Indeed, since $S_{\ell \ell} \varphi \in \mathcal{H}_{-}$ and  $S_{r \ell} \varphi \in \mathcal{H}_{+}$, the scalar products in the integrands converge to zero for each $t \in \mathbb{R}$ as $R \rightarrow \infty$, and it suffices to justify the interchange of the limit and the integration. For this it is enough to majorize the absolute value of the integrands by an $R$-independent integrable function. One has
\begin{eqnarray}\label{Bound Intersection}
&&|\langle e^{-iH_{\ell}t} S_{\ell \ell} \varphi |  \chi_{(-R,R)}
 e^{-iH_{r}t} S_{r \ell} \varphi \rangle| \leq\nonumber\\ 
&&\hspace{1cm}\|\varphi\| \left[\left(\int_{0}^{\infty} |(e^{-iH_{\ell}t} S_{\ell\ell}
  \varphi)(x)|^2 \, dx \right)^{1/2} + \left(\int_{-\infty}^{0} |(e^{-iH_{r}t} S_{r\ell}
  \varphi)(x)|^2 \, dx \right)^{1/2}\right]\p.
\end{eqnarray}
By (\ref{Propagation-b}) and (\ref{Propagation-a}) with $x_0 = 0$ and $\theta > 2$, the \rhs of (\ref{Bound Intersection}) is an integrable function of $t$ on $[0,\infty)$.

To treat the first two integrals in (\ref{Integral 4}) we use the following asymptotic expression from \cite{AC}:
\begin{equation}\label{AS1}
\int_{0}^{\infty} e^{iH_0 t} \chi_{(-R,R)} e^{-iH_0 t} \, dt = \frac{R}{2} H_0^{-1/2} -i
\frac{d}{dE} + {\scriptstyle \cal O}(1) 
\end{equation}
as $R \rightarrow \infty$. This relation holds on $\mathcal{D}^{\mathrm{in}}_2$ and on $\mathcal{D}^{\mathrm{out}}_2$. The derivative $d/dE$ in the second term on the \rhs means differentiation with respect to the energy variable $E$ in $\mathcal{H}^{\mathrm{in}}$ as well as in $\mathcal{H}^{\mathrm{out}}$ (defined by (\ref{Wave In})-(\ref{Wave Out})) \cite{XX2}.

We first consider the terms proportional to $R$ that are obtained when applying (\ref{AS1}) to the first two integrals in (\ref{Integral 4}). Since $S_{\ell \ell}$ commutes with $H_0^{-1/2}$, there is no contribution from the first of these integrals. By using the intertwining relation  $H_{r} S_{r\ell} = S_{r \ell} H_{\ell}$ one obtains the following contribution from the second integral in (\ref{Integral 4}):
\begin{equation}\label{Contribution 2}
 \frac{R}{2}\langle S_{r \ell} \varphi | \{H_0^{-1/2} - (H_0 +V_r -V_\ell)^{-1/2}\} S_{r \ell} \varphi \rangle\p.
\end{equation}
If $V_\ell \not= V_r$ and $S_{r \ell} \varphi \not= 0$, the scalar product in (\ref{Contribution 2}) is strictly positive. Thus, if $V_\ell < V_r$ and $\varphi$ is an incoming state (from the left) which is not entirely reflected by the potential, the local time delay $\tau^{\mathrm{in}}_{R}(\varphi)$ will not admit a finite limit as $R \rightarrow \infty$, more precisely $\tau^{\mathrm{in}}_{R}(\varphi) \rightarrow + \infty$ as  $R \rightarrow \infty$.

Next we observe that the terms from the first two integrals in (\ref{Integral 4}) that are independent of $R$ are just
\begin{eqnarray}\label{TauE}
&& \langle S_{\ell \ell} \varphi | [-i\frac{d}{dE},S_{\ell \ell}] \varphi \rangle + \langle S_{r \ell} \varphi | [-i\frac{d}{dE},S_{r \ell}] \varphi \rangle\nonumber\\
&=& \int_{V_{\ell}}^{\infty} \overline{\varphi^{\mathrm{in}}(E)} \left\{-i\overline{S_{\ell \ell}(E)} \frac{dS_{\ell \ell}(E)}{dE} - i\overline{S_{r \ell}(E)} \frac{dS_{r \ell}(E)}{dE}\right\} \varphi^{\mathrm{in}}(E) \, dE\nonumber\\
&=& \int_{V_{\ell}}^{\infty} \overline{\varphi^{\mathrm{in}}(E)} \mathcal{T}_{\ell \ell}(E) \varphi^{\mathrm{in}}(E) \, dE\p,
\end{eqnarray}
where $\mathcal{T}_{\ell \ell}(E)$ (the expression in the curly bracket) is one of the elements of the Eisenbud-Wigner time delay matrix $\mathcal{T}(E)$ at energy $E$ which will be discussed below.

A similar analysis can be carried through for $\sigma^{\mathrm{out}}_{R}(\varphi)$, defined in (\ref{Sigma Out}), using the asymptotic expression
\begin{equation}\label{AS2}
-\int_{-\infty}^{0} e^{iH_0 t} \chi_{(-R,R)} e^{-iH_0 t} dt = -\frac{R}{2} H_0^{-1/2} -i
\frac{d}{dE} + {\scriptstyle \cal O}(1)\p,
\end{equation}
as $R \rightarrow \infty$. One finds that
\begin{equation}\label{AS3}
\sigma^{\mathrm{out}}_{R}(\varphi) = - \frac{R}{2}\langle S_{r \ell} \varphi | [H_0^{-1/2} - (H_0 +V_r -V_\ell)^{-1/2}] S_{r \ell} \varphi \rangle +  
\int_{V_{\ell}}^{\infty} \overline{\varphi^{\mathrm{in}}(E)} \mathcal{T}_{\ell \ell}(E) \varphi^{\mathrm{in}}(E) \, dE + {\scriptstyle \cal O}(1)\p.
\end{equation}
Again, if $V_\ell \not= V_r$ and if $\varphi$ is not completely reflected, $\tau^{\mathrm{out}}_{R}(\varphi)$ will diverge as $R \rightarrow \infty$, \viz  $\tau^{\mathrm{out}}_{R}(\varphi) \rightarrow - \infty$ as  $R \rightarrow \infty$. However, the divergent term in $\sigma^{\mathrm{out}}_{R}(\varphi)$ is identical with that in 
$\sigma^{\mathrm{in}}_{R}(\varphi)$, except for its sign. Hence the average $\sigma_R(\varphi) = \frac{1}{2} \left[\sigma^{\mathrm{in}}_{R}(\varphi) + \sigma^{\mathrm{out}}_{R}(\varphi)\right]$ converges to a finite limit, given by the \rhs of (\ref{TauE}) and denoted by $\tau(\varphi)$.

The preceding result can be rewritten in terms of the Eisenbud-Wigner time delay operator $\mathcal{T}$. In the representation $\mathcal{H}^{\mathrm{in}}$ of the Hilbert space $\mathcal{H}$ (see (\ref{Wave In})) $\mathcal{T}$ acts at energy $E$ as an operator $\mathcal{T}(E)$ given as follows:\\
(i) for $V_\ell < E < V_r$,  $\mathcal{T}(E)$ is just multiplication by  $\mathcal{T}_{\ell \ell}(E) = -i \overline{S_{\ell \ell}(E)}\, dS_{\ell \ell}(E)/dE$,\\
(ii) for $E > V_r$,  $\mathcal{T}(E)$ acts on $\varphi^{\mathrm{in}}(E)$ as a $2\,\mbox{x}\,2$ matrix, \ie
\begin{equation}\label{TMatrix}
 \mathcal{T}(E) = \begin{pmatrix}
  \mathcal{T}_{\ell \ell}(E) & \mathcal{T}_{\ell r}(E) \\
  \mathcal{T}_{r \ell}(E) & \mathcal{T}_{r r}(E)
\end{pmatrix}\p,
\end{equation}
where (for $a, b = \ell$ or $r$)
\begin{eqnarray}\label{TMatrix2}
\mathcal{T}_{ab}(E) &=& -i \sum_{c = \ell, r} \overline{S_{c a}(E)} \, \frac{d S_{c b}(E)}{dE} = -i \sum_{c = \ell, r} S^*(E)_{a c} \, \frac{d S_{c b}(E)}{dE}\p.
\end{eqnarray}
Since the incoming wave functions considered in this subsection have only one non-zero component ($\varphi^{\mathrm{in}}_{r}(E)=0$), the \rhs of (\ref{TauE}) is just the mean value of the operator $\mathcal{T}$ for the initial wave packet $\varphi$:
\begin{equation}\label{LastFormula}
\tau(\varphi) \equiv \lim_{R \rightarrow \infty} \tau_{R}(\varphi) = \lim_{R \rightarrow \infty} \frac{1}{2} \left[\tau^{\mathrm{in}}_{R}(\varphi) + \tau^{\mathrm{out}}_{R}(\varphi) \right] = \langle \varphi | \mathcal{T} \varphi \rangle\p.
\end{equation}

If $\mu > 2$,  $\mathcal{T}(E)$ is well defined for $E \not = V_\ell, V_r$ (see Appendix~D). Also, as a consequence of the unitarity of $S(E)$, the matrix $\mathcal{T}(E)$ in (\ref{TMatrix}) is hermitian, and $\mathcal{T}_{\ell \ell}(E)$ is real for each $E > V_\ell$ ($E \not = V_r$). This implies that the family $\{\mathcal{T}(E)\}$ determines a unique (in general unbounded) self-adjoint operator which we have denoted by $\mathcal{T}$. This time delay operator $\mathcal{T}$ commutes with $H^{\mathrm{in}}$ and constitutes a quantum-mechanical observable. The next subsection is devoted to some further comments on the meaning of this operator.

To end this subsection we point out an interesting alternative expression for the global time delay $\tau(\varphi)$. We observe that, for $\phi\in\mathcal{H}_+$ as well as for $\phi\in\mathcal{H}_-$ and any $\kappa \in \mathbb{R}$, one has the following identity:
\begin{equation}\label{Invariance Translation Free}
\int_{-\infty}^{\infty} dt \int_{x_1}^{x_2} dx \ |(e^{-iH_{\kappa}t}\phi)(x)|^2 =  \frac{x_2-x_1}{2} \int_{-\infty}^{\infty} |\hat{\phi}(p)|^2 \, \frac{dp}{|p|}\p.
\end{equation}
By using this identity in $T^{\mathrm{in}}_R(\varphi)$ and $T^{\mathrm{out}}_R(\varphi)$ one may write $\tau_R(\varphi)$ as \cite{NN}
\begin{equation}\label{Eq1}
\tau_R(\varphi) = \tau_{R, \ell}(\varphi) + \tau_{R, r}(\varphi) + {\scriptstyle \cal O}(1) \hspace{5mm}\mbox{as} \hspace{5mm} R \rightarrow \infty\p,
\end{equation}
where
\begin{eqnarray*}\label{Eq2}
\tau_{R, \ell}(\varphi) &=& \int_{-\infty}^{\infty} dt \int_{-R}^{0} dx \ \left\{|(e^{-iH t} \Omega^{-}_{\ell} \varphi)(x)|^2 - |(e^{-iH_\ell t} \varphi)(x)|^2 - |(e^{-iH_\ell t} S_{\ell \ell} \varphi)(x)|^2\right\}\p,\\
\tau_{R, r}(\varphi) &=& \int_{-\infty}^{\infty} dt \int_{0}^{R} dx \ \left\{|(e^{-iH t} \Omega^{-}_{\ell} \varphi)(x)|^2 - |(e^{-iH_r t} S_{r \ell} \varphi)(x)|^2\right\}\p.
\end{eqnarray*}
By arguments similar to those applied before \cite{MM} one finds that both $\tau_{R, \ell}(\varphi)$ and $\tau_{R, r}(\varphi)$ converge as $R \rightarrow \infty$, their limits (denoted $\tau_{\ell}(\varphi)$ and $\tau_{r}(\varphi)$ respectively) being given by setting $R=\infty$ in the double integrals defining $\tau_{R, \ell}(\varphi)$ and $\tau_{R, r}(\varphi)$. Thus
\begin{equation}\label{Alternative Exp}
\tau(\varphi) = \tau_{\ell}(\varphi) + \tau_{r}(\varphi)\p.
\end{equation}
This shows that the (symmetrized) global time delay $\tau(\varphi)$ may be decomposed into a contribution associated to the left spatial half-interval $(-\infty,0)$ and a contribution coming from the right half-line $(0,\infty)$. One observes that the final state $S_{\ell}\varphi$ appears in  $\tau_{\ell}(\varphi)$ only through its reflected part $ S_{\ell \ell} \varphi$ and in $\tau_{r}(\varphi)$ only through its transmitted part $S_{r \ell} \varphi$.

\subsection{Discussion}\label{Interpretation}

In the literature and in text books one usually uses the Eisenbud-Wigner expression
\begin{equation}\label{EW Formula}
\langle \varphi | \mathcal{T} \varphi \rangle \equiv \int_{V_{\ell}}^{\infty} \overline{\varphi^{\mathrm{in}}(E)} \left\{-i\overline{S_{\ell \ell}(E)} \frac{dS_{\ell \ell}(E)}{dE} - i\overline{S_{r \ell}(E)} \frac{dS_{r \ell}(E)}{dE}\right\} \varphi^{\mathrm{in}}(E) \, dE
\end{equation}
to compute the time delay induced by a one-dimensional potential $V$ on an incoming wave packet $\varphi \in \mathcal{H}_+$. Its identity with the symmetrized global time delay, Eq.~(\ref{LastFormula}), allows one to have a time-dependent interpretation of what is computed with (\ref{EW Formula}). An alternative interpretation is furnished by the decomposition (\ref{Alternative Exp}) of the symmetrized global time delay.

The symmetrized local time delay $\tau_{R}(\varphi)$ can be written in the following form:
\begin{equation}\label{Tau R Def Alt}
\tau_{R}(\varphi) \equiv \frac{1}{2} [\tau^{\mathrm{in}}_{R}(\varphi) + \tau^{\mathrm{out}}_{R}(\varphi)] = T_{R}(\Omega^{-}_{\ell} \varphi) - \frac{1}{2}
\left[T^{\mathrm{in}}_{R}(\varphi) + T^{\mathrm{out}}_{R}(\varphi)\right]\p.
\end{equation}
We have shown that $\tau_{R}(\varphi)$ is the appropriate
expression admitting a finite limit as $R \rightarrow \infty$.
This implies that the pertinent reference sojourn time $T^{\mathrm{ref}}_{R}(\varphi)$ is
neither $T^{\mathrm{in}}_{R}(\varphi)$ nor $T^{\mathrm{out}}_{R}(\varphi)$ but their
average, \ie
\begin{equation}
T^{\mathrm{ref}}_{R}(\varphi) = \frac{1}{2}
\left[T^{\mathrm{in}}_{R}(\varphi) + T^{\mathrm{out}}_{R}(\varphi)\right]\p.
\end{equation}

Notice that $T^{\mathrm{ref}}_{R}(\varphi)$ depends on the potential $V$ since the free outgoing sojourn time $T^{\mathrm{out}}_{R}(\varphi)$ involves the
final state $S\varphi$. There are however two special cases 
where $T^{\mathrm{ref}}_{R}(\varphi)$ is actually independent of $V$: 

(i) The first one is the situation where $V_{\ell} = V_{r}$. In this case the
difference $T^{\mathrm{in}}_{R}(\varphi) - T^{\mathrm{out}}_{R}(\varphi)$ vanishes
as $R \rightarrow \infty$, so that $T^{\mathrm{ref}}_{R}(\varphi) = T^{\mathrm{in}}_{R}(\varphi) + {\scriptstyle \cal O}(1)$ as $R \rightarrow \infty$. 

(ii) The second one is the general situation $V_{\ell} < V_r$  but with an incoming wave packet $\varphi$ having energy
support (relative to $H_\ell$) contained in the interval $(V_\ell,V_r)$. Since such an incoming state is completely reflected by the scatterer $V$ one has $|(\widehat{S_{\ell \ell} \varphi})(-p)|=|\hat{\varphi}(p)|$ for all $p$, so that $T^{\mathrm{out}}_R(\varphi) = T^{\mathrm{in}}_R(\varphi)$ and therefore $T^{\mathrm{ref}}_{R}(\varphi) = T^{\mathrm{in}}_R(\varphi)$.\\
The statement in (i) is easily obtained by using (\ref{AS1}) and (\ref{AS2}), that in (ii) is a consequence of (\ref{Invariance Translation Free}).

In the above two cases any of the
incoming, outgoing and average reference sojourn time is
acceptable, \ie one has in these cases: 
$$
\tau(\varphi) = \lim_{R \rightarrow \infty} \tau_{R}^{\mathrm{in}}(\varphi) = \lim_{R \rightarrow \infty} \tau_{R}^{\mathrm{out}}(\varphi) = \lim_{R \rightarrow \infty} \tau_{R}(\varphi)\p.
$$
Therefore one can interpret the Eisenbud-Wigner expression (\ref{EW Formula}) as the difference between the sojourn time $T_R(\Omega^{-}_{\ell}\varphi)$ (where $\Omega^{-}_{\ell}\varphi$  evolves with $H=H_0 + V$) and the free incoming sojourn time $T^{\mathrm{in}}_R(\varphi)$ (where $\varphi$  evolves with $H_\ell = H_0 + V_\ell$) as $R \rightarrow \infty$.

We next discuss the situation where $V_{\ell} < V_r$ and $\varphi \in \mathcal{H}_+$ is an incoming wave packet having energy
support (relative to $H_\ell$) above the threshold $V_r$ (\ie in $(V_r,\infty)$). Here the reference sojourn time $T^{\mathrm{ref}}_{R}(\varphi)$ will depend on $V$ (and not just on its asymptotic values $V_\ell$ and $V_r$), so that one cannot have a similar explanation of (\ref{EW Formula}) as the one obtained for the cases (i)-(ii) above. In this case one can still use any of the two expressions given in (\ref{Tau R Def Alt}) or the alternative expression (\ref{Alternative Exp}) as a time-dependent interpretation of $\langle \varphi | \mathcal{T} \varphi \rangle$. Nevertheless it is instructive to introduce the following reference potential: $\mathcal{V} = V_{\ell} \chi_{\ell} + V_{r} \chi_{r}$ (\ie the step-potential represented in dotted lines in Fig.~1). Notice that $\mathcal{V}$ depends on $V$ only through its asymptotic limits $V_\ell$ and $V_r$, and that the point of discontinuity of $\mathcal{V}$ coincides with the center of the interval $[-R,R]$ used to define the local time delay $\tau_{R}(\varphi)$. 

For the step-potential $\mathcal{V}$ one knows explicit expressions for the quantities of interest in scattering theory (see \eg Chapter~1 in \cite{Cohen}). For $E > V_r$ all entries of the $S$-matrix $S(E)$ are real. It then follows from the unitarity relation for $S(E)$ that the diagonal elements (but not the off-diagonal ones) of $\mathcal{T}(E)$ are zero. Therefore the Eisenbud-Wigner expression (\ref{EW Formula}) vanishes if $\varphi \in \mathcal{H}_+$ has energy support (with respect to $H_\ell$) above $V_r$ \cite{XXX}. In this situation (\ref{EW Formula}) is non-zero only if the potential $V$ is different from $\mathcal{V}$. In other words, only the ``wavy'' part $V-\mathcal{V}$ of the potential $V$ may induce a time delay on $\varphi$ and one may therefore interpret the Eisenbud-Wigner expression as the effect of $V-\mathcal{V}$. Note that, although the scatterer $\mathcal{V}$ induces no time delay above $V_r$, it still affects the incoming wave packet $\varphi$ ($S \not = I$ if $H=H_0 + \mathcal{V}$). Note also that any potential $V$ with $V_\ell < V_r$ induces an infinite incoming global time delay above $V_r$ ($\tau^{\mathrm{in}}(\varphi)=+\infty$) so that the Eisenbud-Wigner formula (or equivalently the symmetrized global time delay) is the appropriate expression to measure the finite effect of the wavy part $V-\mathcal{V}$ of the potential $V$.

In conclusion, for a state $\varphi$ incident from the left, one may interpret the symmetrized global time delay, or equivalently the Eisenbud-Wigner expression, as the effect of the
full potential $V$ on
the components $\varphi^{\mathrm{in}}(E)$ of $\varphi$ with energy $E$
below $V_r$ and as the effect of the wavy part $V-\mathcal{V}$ of $V$ on
the components $\varphi^{\mathrm{in}}(E)$ with energy above $V_r$. 

\section{Concluding Remarks}\label{Concluding Remarks}

We have presented a general scattering theory and obtained the existence of the global time delay as well as its identity with the Eisenbud-Wigner expression for potentials $V$ having different limits at $x = -\infty$ and at $x = +\infty$, assuming $V$ to be bounded and approaching its limits at $x = \pm\infty$ at a certain minimal rate (specified by the number $\mu$ in (\ref{Assumption VL})-(\ref{Assumption VR})). From this identity we obtained a time-dependent interpretation of the Eisenbud-Wigner expression. The same results can be established under weaker assumptions on $V$ by using more refined versions of our approach or different techniques. It is possible to handle potentials with (square-integrable) local singularities and to weaken the assumptions that we made on $\mu$. The hypothesis $\mu > 1$ (short range condition) is sufficient for obtaining all results on scattering theory in Section~\ref{Scattering theory} \cite{TT}. As regards time delay we expect (in view of known results \cite{XX} for $n$-dimensional Hamiltonians with potentials converging to zero as $|x| \rightarrow \infty$) that the symmetrized global time delay should exist and coincide with the Eisenbud-Wigner expression if $\mu > 2$ and for initial states belonging to $\mathcal{D}^{\mathrm{in}}_{\theta}$ for some $\theta > 2$.

Only wave functions with energy support away from the thresholds $V_\ell$ and $V_r$ have been considered. At thresholds the time delay is usually infinite. As a consequence the Eisenbud-Wigner operator $\mathcal{T}$ will be unbounded, and its behavior near the thresholds would require a more refined investigation.

In one-dimensional scattering problems it is interesting to distinguish between the reflected and the transmitted part of a wave function, and there is a considerable literature on tunneling times (\cite{HS}, \cite{NHM} and references cited therein). The value of the global time delay for a given initial state involves both the reflected and the transmitted wave and is therefore not related directly to a tunneling time. However the global time delay has the merit of being given in terms of a self-adjoint linear operator and can thus be interpreted as a quantum-mechanical observable in the usual sense; furthermore it has a meaning also for scattering systems in more than one space dimension.

We point out that no oscillating terms in $R$ were involved when we considered the limit of the local time delays as $R \rightarrow \infty$. In various other publications on time delay (\eg \cite{Smith}, \cite{JM}, \cite{BO}, \cite{CN}) the authors encountered oscillatory terms like $\sin(2 p R)$ and then presented an argument to suggest that these terms will not contribute to the global time delay. In \cite{CN} it is stated that these oscillatory terms are related to the uncertainty principle. Actually, the presence of such terms arises when one works with non-normalizable eigenfunctions of the Hamiltonian. In a fully Hilbert space derivation of global time delay, with square-integrable wave functions, one has no problem with oscillatory terms (see also \cite{BO}).

One could also consider the global time delay $\tau^{x_{0}}(\varphi)$ obtained by starting with the local time delay $\tau^{x_{0}}_{R}(\varphi)$ in the translated interval $[-R+x_{0}, R+x_{0}]$, where $x_{0} \in \mathbb{R}$. It is clear that, for $\varphi \in \mathcal{H}_{+}$, the local time delay $\tau^{x_{0}}_{R}(\varphi)$ is given by (\ref{Time-Delay-Average}) with the following substitutions: $\varphi \mapsto \varphi^{x_{0}} = e^{iPx_{0}} \varphi$ and $S \mapsto S^{x_{0}} = e^{iPx_{0}} S  e^{-iPx_{0}}$. Then, proceeding as in \cite{GT} or \cite{SM}, one obtains
\begin{equation}\label{Tauc}
\tau^{x_{0}}(\varphi) = \tau(\varphi) - \frac{x_{0}}{2} \langle S_{\ell \ell} \varphi |  [P^{-1},  S_{\ell \ell}]  \varphi \rangle -  \frac{x_{0}}{2} \langle S_{r \ell} \varphi | [P^{-1}, S_{r \ell}]  \varphi \rangle\p.
\end{equation}
Thus two additional terms appear in the above situation. The presence of such terms was already pointed out in the case $V_\ell = V_r$ \cite{JW1} and also in a more general situation \cite{GT}. Recalling the identity (\ref{LastFormula}) one sees from (\ref{Tauc}) that the Eisenbud-Wigner expression for time delay assumes implicitly that the spatial interval $[-R+x_0,R+x_0]$ on which the total and reference sojourn times are compared is centered at the origin ($x_0 = 0$). Note in particular that the relation (\ref{Tauc}) implies that the time delay $\tau(\varphi)$ due to the translated step-potential $\mathcal{V} = V_{\ell} \, \chi_{(-\infty,x_0)} + V_{r} \, \chi_{(x_0,\infty)}$ ($x_0 \not = 0$) for a wave packet $\varphi \in \mathcal{H}_+$ having energy support (with respect to $H_\ell$) above $V_r$ is non-zero in general. Finally, notice that the sojourn times (\ref{Sojourn-Time2})-(\ref{Sojourn-Time-Out2}) and therefore the symmetrized time delay $\tau(\varphi)$ are invariant under time translations. 

As a last point we mention that one can find a general study and discussion of the symmetrized definition of time delay in \cite{GT}. These authors point out in particular that the symmetrized time delay is useful in multichannel scattering processes, invariant under an appropriate mapping of time reversal and relatively insensitive to the shape of the spatial regions used for defining the local time delay in more than one space dimension.

\section*{ACKNOWLEDGMENTS}\label{ACKNOWLEDGMENTS}

This investigation was suggested to us by S.~Richard. We also have benefited from numerous discussions with him and with V.~Cachia and R.~Tiedra de Aldecoa to all of whom we wish to express our sincere gratitude. This work was partially supported by the Swiss National Science Foundation.

\renewcommand{\theequation}{A-\arabic{equation}}
\setcounter{equation}{0} 
\section*{APPENDIX A: TIME DECAY IN A CONSTANT POTENTIAL}

We show here how to obtain (\ref{Propagation-a}) and (\ref{Propagation-c}). The estimates (\ref{Propagation-a2})-(\ref{Propagation-b2})
 can be deduced in a similar manner. We assume (without loss of generality) that $x_{0}=0$ in (\ref{Propagation-a}). We first consider the case $|t|\leq 1$ and let $\varphi$ be any wave packet. If $g$ is a bounded function, say $|g(x)| \leq M < \infty$, then for $|t|\leq 1$ and $\theta \geq 0$:
\begin{eqnarray*}
\int_{-\infty}^{\infty} |g(x) \varphi_{t}(x)|^2 \, dx &\leq& M^2 \int_{-\infty}^{\infty} |\varphi(x)|^2 \, dx\\
&\leq& \frac{2^{\theta} M^2}{(1+|t|)^{\theta}} \int_{-\infty}^{\infty}
|\varphi(x)|^2 \, dx\p.
\end{eqnarray*}
Taking $g$ satisfying $g(x)=1$ if $x \leq 0$ and $g(x)=0$ otherwise, or $g=f$ with $f$ as in (\ref{Propagation-c}), one sees that the inequalities (\ref{Propagation-a}) and (\ref{Propagation-c}) are satisfied for $|t|\leq 1$.

To treat the remaining values of t ($t > 1$ in (\ref{Propagation-a}) and $|t| > 1$ in (\ref{Propagation-c})), we write $\theta$ in the form $\theta = 2n + 2 \epsilon$, with $n$ a non-negative integer and $0 \leq \epsilon < 1$, and we shall use Taylor's formula for the
function $F(s) = e^{is}$ ($s\in \mathbb{R}$), \ie
\begin{equation}\label{A3}
e^{is} = \sum_{k=0}^{n} \frac{i^k s^k}{k!} + R_{n}(s)
\end{equation}
with
\begin{equation}\label{A4}
|R_{n}(s)| \leq (1+e) |s|^{n+\epsilon} \leq 4 |s|^{n+\epsilon}
\end{equation}
for $0 \leq \epsilon \leq 1$.

Let $\varphi$ be a wave packet with positive momentum and let $g =
1 - h$, where $h$ is the Heaviside function (so $g(x) = 0$ for $x
> 0$ and $g(x) = 1$ for $x < 0$). Let $t \geq 1$. Then $g(2tp) = 0$
for all $p > 0$, so that $g(2tp)\hat{\varphi}(p)\equiv 0$. Also,
since $Q$ is just differentiation in momentum space, the wave
function $Q^{k}\varphi$ has the same momentum support as $\varphi$
for any positive integer k (assuming $\hat{\varphi}$ at least $k$
times differentiable). Hence $g(2tp) (\mathcal{F}Q^{k}\varphi)(p)
\equiv i^{k}g(2tp)\hat{\varphi}^{(k)}(p) = 0$ for all real $p$. Thus, taking for example $n=1$ (\ie $2 \leq \theta < 4$), we get from (\ref{Identity x-p}) that
\begin{eqnarray*}
\int_{-\infty}^{0} |\varphi_{t}(x)|^2 \, dx &=& \int_{-\infty}^{\infty} |g(2tp) [\mathcal{F}(e^{iQ^2/4t}\varphi)](p)|^2 \, dp\\
&=&  \int_{-\infty}^{\infty} |g(2tp) [\mathcal{F}(e^{iQ^2/4t}\varphi - \varphi - i\frac{Q^2}{4t}\varphi)](p)|^2 \, dp\p.
\end{eqnarray*}
By using the bound $|g(2tp)|\leq 1$, the unitarity of
$\mathcal{F}$ and (\ref{A3})-(\ref{A4}) with $n = 1$, one
obtains that
\begin{eqnarray*}
\int_{-\infty}^{0} |\varphi_{t}(x)|^2 \, dx &\leq&  \int_{-\infty}^{\infty} |(e^{ix^2/4t} - 1 - i\frac{x^2}{4t})\varphi(x)|^2 \, dx\\
&\leq&  \frac{1}{|t|^{2(1+\epsilon)}} \int_{-\infty}^{\infty} | |x|^{2(1+\epsilon)}\varphi(x)|^2 \, dx
\end{eqnarray*}
which implies the validity of (\ref{Propagation-a}) for $t > 1$ and $\theta \in [2,4)$. The result for other values of $\theta$ is obtained similarly (using an $n$-th order Taylor expansion if $\theta = 2(n+\epsilon)$).

We finally show how to verify (\ref{Propagation-c}) for $|t| > 1$ and $\theta \in [4,6)$. Let $f$ be as stated in (c). Using
the inequality $|\alpha + \beta|^{2}\leq 2 |\alpha|^2 + 2
|\beta|^2$ we get from  (\ref{Identity x-p}) that
\begin{eqnarray}\label{A6}
\int_{-\infty}^{\infty} |f(x) \varphi_{t}(x)|^2 \, dx &=&  \int_{-\infty}^{\infty} |f(2tp)|^2 |[\mathcal{F}(e^{iQ^2/4t}\varphi)](p)|^2 \, dp\nonumber\\
&\leq&  2 \int_{-\infty}^{\infty}  |f(2tp)|^2 |[\mathcal{F}(e^{iQ^2/4t}\varphi - \varphi -i\frac{Q^2}{4t}\varphi + \frac{Q^4}{32t^2}\varphi)](p)|^2 \, dp\nonumber\\
&+&  2 \int_{-\infty}^{\infty}  |f(2tp)|^2 |[\mathcal{F}(\varphi
+i\frac{Q^2}{4t}\varphi - \frac{Q^4}{32t^2}\varphi)](p)|^2 \, dp\p.
\end{eqnarray}
In the first integral on the \rhs of (\ref{A6}) we majorize
$|f(2tp)|^{2}$ by $C^{2}$ and use Taylor's formula (with
$n = 2$) to obtain, as above, the following upper bound for this
term:
\begin{equation*}
2 C^2 \frac{1}{|t|^{2(2+\epsilon)}} \int_{-\infty}^{\infty} | |x|^{2(2+\epsilon)} \varphi(x)|^2 \, dx
\end{equation*}
which is majorized by the first contribution in (\ref{Propagation-c}), with $\theta = 2(2+\epsilon)$ (and $C_{\theta}=2^{\theta+1} C^2$). To treat
the second term on the \rhs of (\ref{A6}) we set $\eta = \varphi
+ iQ^{2}\varphi/4t - Q^{4}\varphi/32t^{2}$ and observe that
$\hat{\eta}(p)\not= 0$ only for $|p|\geq p_{0}$ if $\varphi$ has
momentum in $\mathbb{R}\setminus (-p_{0},p_{0})$. We majorize $|f(2tp)|$ by
$C (1+2 |p_{0}| |t|)^{-\mu}$ in the integral and get
\begin{eqnarray*}
2 \int_{-\infty}^{\infty} |f(2tp)|^2 |\hat{\eta}(p)|^2 \, dp &\leq&  \frac{2 C^2}{(1+2p_{0}|t|)^{2\mu}} \int_{-\infty}^{\infty} |\hat{\eta}(p)|^2 \, dp\\
&\leq&  \frac{2 C^2}{(1+2p_{0}|t|)^{2\mu}} \int_{-\infty}^{\infty}
|\eta(x)|^2 \, dx
\end{eqnarray*}
which is majorized by the second contribution in (\ref{Propagation-c}) if $|t| > 1$. 

\renewcommand{\theequation}{B-\arabic{equation}}
\setcounter{equation}{0} 
\section*{APPENDIX B: THE MOURRE ESTIMATE}

We establish here the validity of a Mourre estimate, as stated in
Section~\ref{Time decay in a non-constant potential}, for our class of Hamiltonians ($\mu > 1$). We shall freely use the following properties of compact
operators. The product of a compact operator and a bounded
operator is compact. If $\{K_{n}\}$ is a sequence of compact
operators that converges in operator norm, \ie if there exists a
bounded operator $K$ such that $\|K_{n} - K\|_{\mathcal{B}(\mathcal{H})} \rightarrow 0$ as
$n\rightarrow \infty$, then $K$ is also compact. If f is a bounded
function on $\mathbb{R}$ satisfying $f(x) = 0$ near $x = \pm
\infty$ or, more generally, such that $f(x) \rightarrow 0$ as $|x|
\rightarrow \infty$ then, for our
class of Hamiltonians, the operator $(H-z)^{-1}f(Q)$ is compact
for each non-real~$z$. Also, if $F(\Delta)$ denotes the projection
onto the subspace of wave functions having energy support (with respect to $H$) in the
interval $\Delta$, then $f(Q)F(\Delta)$ and  $f(Q) P F(\Delta)$ are compact if $f$ is as
above and $\Delta$ is a bounded interval.

We fix a smooth function $J_{r}$ satisfying $0 \leq J_{r}(x) \leq 1$ for all $x$, $J_{r}(x) = 0$ for $x \leq 0$ and $J_{r}(x) = 1$ for $x \geq 1$,
 and we set
$J_{\ell} = 1-J_{r}$. We denote also by $J_{r}$ the operator of
multiplication by $J_{r}(x)$. The following decomposition of $V$ will be used:
\begin{equation}\label{Decomposition V}
V = (V-V_{\ell}) J_{\ell} + (V-V_{r})J_{r} + V_{\ell} J_{\ell} + V_{r} J_{r}\p.
\end{equation}
We consider wave packets $\psi$ with
energy support in a finite interval $\Delta = [\alpha,\beta]$, \ie
satisfying $\psi = F(\Delta)\psi$. For $A = (PQ+QP)/4$, one has
$[iH_{0},A] = [iP^{2},A] = P^{2} = H-V$ and $[iV,A] = -QV'/2$ (assuming that $V$ is differentiable, see the Remark below). Hence
\begin{eqnarray}\label{A7}
\langle \psi |[iH,A] \psi \rangle &=&  \langle \psi |H \psi \rangle - \langle \psi |V \psi \rangle -  \frac{1}{2} \langle \psi |QV' \psi \rangle\nonumber\\
 &\geq& \alpha \langle \psi | \psi \rangle - V_{\ell} \langle \psi | J_{\ell}\psi \rangle - V_{r} \langle \psi | J_{r}\psi \rangle\nonumber\\
& &  - \langle \psi |(V - V_{\ell}) J_{\ell} \psi \rangle - \langle \psi |(V - V_{r}) J_{r} \psi \rangle - \frac{1}{2}\langle \psi |QV' \psi \rangle\p.
\end{eqnarray}

(i) We first assume that $V_{r} < \alpha < \beta < \infty$. One has
\begin{equation*}
 V_{\ell} \langle \psi | J_{\ell}\psi \rangle +  V_{r} \langle \psi | J_{r}\psi \rangle \leq V_{r}  \langle \psi |(J_{\ell} + J_{r})\psi \rangle =  V_{r}  \langle \psi | \psi \rangle\p.
\end{equation*}
Hence (\ref{A7}) leads to
\begin{equation*}
 \langle \psi |[iH,A]\psi \rangle \geq (\alpha-V_{r})  \langle \psi | \psi \rangle - \langle \psi |[(V-V_{\ell})J_{\ell} + (V-V_{r})J_{r} + \frac{1}{2}QV'] \psi
 \rangle\p.
\end{equation*}
Since $(V(x)-V_{\ell})J_{\ell}(x) + (V(x)-V_{r})J_{r}(x) + \frac{1}{2}xV'(x)$  converges to zero
as $x \rightarrow\pm \infty$ (provided that $V'$ decays more rapidly than $|x|^{-1}$), the Mourre inequality (\ref{Mourre Estimate}) holds
with $\lambda = \alpha-V_{r}$ and $K = -F(\Delta) [(V-V_{\ell})J_{\ell} + (V-V_{r})J_{r} + \frac{1}{2}QV'] F(\Delta)$.

\textbf{Remark.} If $V$ is not differentiable the compactness of $F(\Delta) [iV,A] F(\Delta)$ is obtained (for $\mu > 1$) by decomposing $V$ as in (\ref{Decomposition V}) and writing for example $2[iV_{\ell}J_{\ell},A]=-V_{\ell}QJ'_{\ell}$ and $2[i(V-V_{\ell})J_{\ell},A]=i[Q(V-V_{\ell})J_{\ell}]P -i P[Q(V-V_{\ell})J_{\ell}] + (V-V_{\ell})J_{\ell}$.

(ii) We now consider the case where $V_{\ell} < \alpha < \beta < V_{r}$. Since we assumed that $\psi = F(\Delta) \psi$, the sum of the last three terms in (\ref{A7}) is again of
the form $\langle \psi | K_{0}\psi \rangle$ for some compact
operator $K_{0}$. We shall show that, for $\Delta \subset
(V_{\ell},V_{r})$, the operator $J_{r}F(\Delta)$ is compact. Writing $V_{\ell} \langle \psi | J_{\ell} \psi \rangle = V_{\ell} \langle \psi | \psi \rangle - V_{\ell} \langle \psi | J_{r} \psi \rangle$, it then follows from (\ref{A7}) that a Mourre estimate holds with $\lambda = \alpha - V_{\ell}$ and $K = K_{0} + F(\Delta)(V_{\ell} - V_{r})J_{r}F(\Delta)$.

Let $g$ be a smooth function defined on $\mathbb{R}$ such that
$g(E) = 1$ for $E \in \Delta$, $g(E) = 0$ for $E>(\beta+V_{r})/2$ and
for $E<V_{\ell}$. If $\psi$ has energy support in $\Delta$, then $\psi =
g(H)\psi$, which implies that $g(H)F(\Delta) = F(\Delta)$, and it
suffices to show that $J_{r}g(H)$ is a compact operator. For this
we introduce an auxiliary Hamiltonian $\hat{H} = H_{0} + \hat{V}$,
with $\hat{V}(x) = V(x)$ for $x \geq 0$ and $\hat{V}(x) = V_{r}$ for $x
< 0$. We may write
\begin{equation}\label{A9}
J_{r}g(H) = J_{r}g(\hat{H}) + J_{r}[g(H) - g(\hat{H})]\p.
\end{equation}
The operator  $g(\hat{H})$ (hence also $J_{r}  g(\hat{H})$) is
compact (even of finite rank), because the Hamiltonian $\hat{H}$
has only discrete spectrum below its threshold $V_{r}$, hence at most a finite number of eigenvalues below $(\beta+V_{r})/2$.

To handle the second term on the \rhs of (\ref{A9}), we use the
following formula for $g(H)$ (Theorem 6.1.4 and Remark 6.1.3 of \cite{ABG}):
\begin{eqnarray}\label{A10}
g(H) &=& \frac{1}{2\pi i} \int_{-\infty}^{\infty} g(E) \, [(H-E-i)^{-1} - (H-E+i)^{-1}] \, dE\nonumber\\
&+& \frac{1}{2\pi} \int_{-\infty}^{\infty} g'(E) \, [(H-E-i)^{-1} + (H-E+i)^{-1}] \, dE\nonumber\\
&-& \frac{1}{2\pi i} \int_{0}^{1} \epsilon \, d\epsilon
\int_{-\infty}^{\infty} g''(E)\, [(H-E-i\epsilon)^{-1} -
(H-E+i\epsilon)^{-1}] \, dE\p.
\end{eqnarray}
The integrals in (\ref{A10}) exist in operator norm (\ie the
approximating Riemann sums converge in operator norm). By using
(\ref{A10}) and the corresponding formula for $g(\hat{H})$, one
has (with the notations $R_{z} = (H-z)^{-1}$ and $\hat{R}_{z} =
(\hat{H}-z)^{-1}$):
\begin{eqnarray}\label{A11}
J_{r} [g(H)-g(\hat{H})] &=& \frac{1}{2\pi i} \int_{-\infty}^{\infty} g(E) \, [J_{r} (R_{E+i} - \hat{R}_{E+i}) - J_{r} (R_{E-i} - \hat{R}_{E-i})] \, dE\nonumber\\
&+& \frac{1}{2\pi} \int_{-\infty}^{\infty} g'(E) \, [J_{r} (R_{E+i} - \hat{R}_{E+i}) + J_{r} (R_{E-i} - \hat{R}_{E-i})] \, dE\nonumber\\
& & \hspace{-20mm} - \frac{1}{2\pi i} \int_{0}^{1} \epsilon \, d\epsilon
\int_{-\infty}^{\infty} g''(E) [J_{r} (R_{E+i\epsilon} -
\hat{R}_{E+i\epsilon}) - J_{r} (R_{E-i\epsilon} -
\hat{R}_{E-i\epsilon}) ] \, dE\p.
\end{eqnarray}
Since the integrals in (\ref{A11}) exist in norm, this implies the
compactness of $J_{r}[g(H) - g(\hat{H})]$ by using the fact
(established below) that $J_{r}(R_{z} -\hat{R}_{z})$ is a compact
operator for each non-real~$z$.

To verify the compactness of $J_{r}(R_{z} - \hat{R}_{z})$, we write
\begin{eqnarray}
J_{r}(R_{z} - \hat{R}_{z}) &\equiv& J_{r}[(H-z)^{-1} - (\hat{H}-z)^{-1}]\nonumber\\
&=& J_{r}(H-z)^{-1}(\hat{V}-V)(\hat{H}-z)^{-1}\nonumber\\
&=& (H-z)^{-1}J_{r}(\hat{V}-V)(\hat{H}-z)^{-1} + \nonumber\\
& & + (H-z)^{-1}[H-z,J_{r}](H-z)^{-1}(\hat{V}-V)(\hat{H}-z)^{-1}\p.\label{A12}
\end{eqnarray}
The first term on the \rhs of (\ref{A12}) is zero since $J_{r}(x)[\hat{V}(x)-V(x)] \equiv 0$. In the second
term we observe that $[H-z,J_{r}] = [P^{2},J_{r}] = -J_{r}'' -
2iJ_{r}'P$, and the compactness of the second term follows since
$J_{r}''(x) = J_{r}'(x) = 0$ for $|x| \geq 1$ and $P(H-z)^{-1}$ is
a bounded operator.

\renewcommand{\theequation}{C-\arabic{equation}}
\setcounter{equation}{0} 
\section*{APPENDIX C: EXISTENCE OF LARGE TIME LIMITS}

In this appendix we prove the existence of the (strong) limits involved in the definitions of the scattering projections $F^{\pm}_{\ell}$ and $F^{\pm}_{r}$ (see (\ref{Projectors-L})-(\ref{Projectors-R})) and of the operator $\mathcal{W}$ given by Eq.~(\ref{Separating LR3}). It suffices to establish the existence of these limits on a dense set of wave functions in $\mathcal{H}_{c}(H)$. We consider the dense set of $\psi$ having bounded energy support (with respect to $H$) away from the thresholds $V_{\ell}$ and $V_{r}$. 

(a) We first consider $F^{+}_{\ell}$, as given by (\ref{F+L}), for which we use (\ref{W Cauchy}) with $h_{1}=h_{2}=H$, hence $W_{\tau}=e^{iH\tau} g(Q) e^{-iH\tau}$. Due to the first projection $F_{c}(H)$ in (\ref{F+L}), the supremum over the set $\{\phi \in \mathcal{H} \ | \ \|\phi\|=1 \}$ in (\ref{W Cauchy}) can be replaced by that over the set $\{\phi \in \mathcal{H}_{c}(H) \ | \ \|\phi\|=1 \}$. By the Cauchy-Schwarz inequality the integrand in (\ref{W Cauchy}), with $V_1=V_2$, is majorized by
\begin{equation}\label{Bound 1+Q}
\|(1+|Q|)^{-1} e^{-iH\tau} \phi\| \cdot \|[(1+|Q|)g''(Q) + 2i(1+|Q|)g'(Q)P]  e^{-iH\tau} \psi\|\p.
\end{equation}
Inserting this bound into (\ref{W Cauchy}) and then applying the Schwarz inequality in $L^2([s,t])$ (\ie to the integral over the interval $[s,t]$), one has
\begin{equation}\label{W Exists}
\|W_{t}\psi - W_{s}\psi\|^2 \leq \sup_{\phi\in\mathcal{H}_{c}(H), \|\phi\|=1} \int_{s}^{t} d\omega \int_{-\infty}^{\infty} dx \ |[(1+|Q|)^{-1} e^{-iH\omega}\phi](x)|^2 \cdot \int_{s}^{t} N_{\tau}^2 \, d\tau\p,
\end{equation}
where $N_{\tau}$ is defined as the second factor in
(\ref{Bound 1+Q}). Let $\Sigma$ be a bounded closed set in $(V_\ell,\infty)$
disjoint from $V_{r}$. By
(\ref{Propagation-Estimate-Strong-Consequence}) there is a
constant $C_{\Sigma}$ such that
$$
\int_{-\infty}^{\infty} d\omega \int_{-\infty}^{\infty} dx \ |[(1+|Q|)^{-1}
e^{-iH\omega}\phi](x)|^2 \leq C_{\Sigma} \ \|\phi\|^2
$$
for all $\phi$ in $\mathcal{H}_{c}(H)$ having energy support in $\Sigma$. By restricting the
supremum in (\ref{W Exists}) to this set of $\phi$, one finds
that
\begin{equation}
\|F(\Sigma)W_{t}\psi - F(\Sigma)W_{s}\psi\|^2 \leq C_{\Sigma} \int_{s}^{t} N_{\tau}^2 \, d\tau\p,
\end{equation}
where $F(\Sigma)$ denotes the projection onto the subspace of wave
functions having energy support (with respect to $H$) in $\Sigma$. We show below that $\int_{0}^{\infty}
N_{\tau}^2 \, d\tau < \infty$. Hence the sequence $\{F(\Sigma)W_{\tau}\psi\}$ is Cauchy as $\tau \rightarrow +\infty$ and therefore 
$\mbox{s-lim}_{t \rightarrow +\infty} F(\Sigma)
e^{iHt}g(Q)e^{-iHt}F_{c}(H)$ exists. By varying the set $\Sigma$,
one can conclude that $\mbox{s-lim}_{t \rightarrow +\infty}
F_{c}(H) e^{iHt}g(Q)e^{-iHt}F_{c}(H)$ exists \cite{Comment-3}.

We now comment on the finiteness of $\int_{0}^{\infty} N_{\tau}^2 \, d\tau$. For the first term in $N_{\tau}$ (\ie $\|(1+|Q|) g''(Q)  e^{-iH\tau} \psi\|$) this is immediate from (\ref{Propagation-Estimate-Strong-Consequence}) since $g''(x)=0$ outside the interval $[-1,0]$, hence $(1+|x|) g''(x) \leq C (1+|x|)^{-1}$. The second term in $N_{\tau}$ requires more care because $P$ is an unbounded operator and does not commute with $e^{-iHt}$. By a commutator calculation (see (c) below) this term can be expressed as a sum of four terms of the form $D f(Q) (H+i)^{m}\psi_{\tau}$, with $D$ a bounded operator, $f$ satisfying $|f(x)| \leq C (1+|x|)^{-1}$ and $m=0$ or $m=1$. Clearly
\begin{eqnarray}\label{Eq Com Form}
& & \int_{-\infty}^{\infty} d\tau \int_{-\infty}^{\infty} dx \ |[D f(Q) (H+i)^{m}\psi_{\tau}](x)|^2 \nonumber\\
&\leq& C^2 \|D\|_{\mathcal{B}(\mathcal{H})}^2 \int_{-\infty}^{\infty} \|(1+|Q|)^{-1} e^{-iH\tau} (H+i)^{m}\psi]\|^2  \, d\tau < \infty
\end{eqnarray}
since $(H+i) \psi$ belongs to $\mathcal{H}_{c}(H)$ and has the same (bounded) energy support as $\psi$.

(b) The arguments for the existence of $\mbox{s-lim}_{t\rightarrow +\infty} e^{iH_{\ell} t} g(Q) e^{-iH t} F_c(H)$ (implying that of $\mathcal{W}$ in (\ref{Separating LR3})) are very similar to those used in (a) above. The only difference arises through the fact that now $h_1=H_\ell$ (instead of $h_1=H$). This leads to an additional term in $\int_{s}^{t} N^2_{\tau} \, d\tau$, \viz
\begin{equation}\label{Finite2}
\int_{s}^{t} d\tau \int_{-\infty}^{\infty} dx \ |[(1+|Q|) (V-V_{\ell}) g(Q) e^{-iH\tau} \psi](x)|^2\p.
\end{equation}
Again it suffices to know that the integral with respect to $d\tau$ in (\ref{Finite2}) is finite if $t=+\infty$; this follows from (\ref{Propagation-Estimate-Strong-Consequence})  since $|(1+|x|) [V(x)-V_{\ell}] g(x)| \leq C (1+|x|)^{-1}$ if $\mu \geq 2$ in (\ref{Assumption VL}).

(c) In (a) above we used the following formula, with $\phi = (1+|x|) g'$ a smooth function vanishing outside some finite interval $\Delta$:
\begin{equation}\label{C1}
\phi(Q) P = \sum_{k=1}^{4} D_{k} f_{k}(Q) (H+i)^{m_k}
\end{equation}
with $D_{k}$ bounded operators, $f_{k}$ smooth functions vanishing outside $\Delta$ and $m_{k}=0$ or $1$. This can be obtain as follows:
\begin{eqnarray*}
\phi(Q)P &=& \phi(Q)P(H+i)^{-1}(H+i)\\
&=& P(H+i)^{-1}\phi(Q)(H+i) + [\phi(Q),P(H+i)^{-1}](H+i)\p.
\end{eqnarray*}
Evaluation of the commutator gives
\begin{eqnarray*}
[\phi(Q),P(H+i)^{-1}] &=& [\phi(Q),P](H+i)^{-1} + P [\phi(Q),(H+i)^{-1}]\\
&=& i\phi'(Q)(H+i)^{-1} - P(H+i)^{-1} [\phi(Q),H+i](H+i)^{-1}\\
&=& i\phi'(Q)(H+i)^{-1} - P(H+i)^{-1} \{-\phi''(Q) + 2iP\phi'(Q)\} (H+i)^{-1}\p.
\end{eqnarray*}
So
\begin{eqnarray*}
\phi(Q)P &=& P(H+i)^{-1}\phi(Q)(H+i) +  i\phi'(Q)\\
&& + P(H+i)^{-1}\phi''(Q) - 2i P(H+i)^{-1}P\phi'(Q)\p,
\end{eqnarray*}
which is of the form (\ref{C1}) since $P(H+i)^{-1}$ and $P(H+i)^{-1}P$ are bounded operators.

\renewcommand{\theequation}{D-\arabic{equation}}
\setcounter{equation}{0} 
\section*{APPENDIX D: DIFFERENTIABILITY OF THE S-MATRIX} 

The entries of the $S$-matrix $S(E)$, \ie the transmission and reflection amplitudes at energy $E$, are given as simple functions of $E$ combined with Wronskians of Jost solutions for the potential $V$. We refer to \cite{Gesztesy} or \cite{Po} for a complete description of these expressions and to Section~XVII.1 of \cite{CS} for a presentation of their derivation. For example the transmission amplitude $S_{r\ell}(E)$ at energy $E > V_r$ is just
\begin{equation}
S_{r\ell}(E) = \frac{2i(k_\ell k_r)^{1/2}}{W(f_\ell(E),f_r(E))}\p,
\end{equation}
where $k_\ell = (E-V_\ell)^{1/2}$,  $k_r = (E-V_r)^{1/2}$ and the Wronskian of the left and right Jost solutions $f_{\ell}$ and $f_r$ is ($' = \partial/\partial x$)
\begin{equation*}
W(f_\ell(E),f_r(E)) = f_\ell(E,x) f'_r(E,x) - f_r(E,x) f'_\ell(E,x)\p. 
\end{equation*}
Thus differentiability or H\"{o}lder continuity of the Jost solutions and their spatial derivatives, as functions of $E$, imply differentiability or H\"{o}lder continuity of $S_{r\ell}(E)$.

Properties of this type can be obtained from the Volterra integral equations for $f_\ell$ and $f_r$. Consider the following integral equation for a function $\Psi_{\kappa,\sigma}(x)$, where $\kappa$ and $\sigma$ are two parameters:
\begin{equation}\label{(A)}
\Psi_{\kappa,\sigma}(x) = \Phi(\kappa,\sigma;x) + \int_{x}^{\infty} \frac{\sin\kappa(y-x)}{\kappa} \mathcal{V}(y) \Psi_{\kappa,\sigma}(y) \, dy\p,
\end{equation}
assuming that the inhomogeneous term $\Phi(\kappa,\sigma;x)$ satisfies an inequality of the form
\begin{equation}\label{(B)}
|\Phi(\kappa,\sigma;x)| \leq \eta(\kappa,\sigma) h(x)
\end{equation}
with $h(x) \geq 1$ and 
\begin{equation}\label{(C)}
\int_{0}^{\infty} h(y) |\mathcal{V}(y)| \, dy < \infty\p.
\end{equation}
The standard iterative method for solving Volterra equations gives the following bound on the solution $\Psi_{\kappa,\sigma}(x)$ for $x \geq 0$:
\begin{equation}\label{(D)}
|\Psi_{\kappa,\sigma}(x)| \leq \eta(\kappa,\sigma) h(x) \exp\left(\frac{1}{\kappa} \int_{0}^{\infty} h(y) |\mathcal{V}(y)| \, dy\right)\p.
\end{equation}

The integral equation for the Jost solution $f_r$ is of the above form, with $\Psi_{\kappa,\sigma}(x) \equiv \Psi_{\kappa}(x) = f_{r}(E,x)$, $\kappa = k_r=(E-V_r)^{1/2}$, $\mathcal{V}(x)=V(x)-V_r$ and $\Phi(\kappa,\sigma;x)=e^{i\kappa x}$, hence $\eta(\kappa,\sigma)=1$ and $h\equiv 1$. The existence of $f_r$ (for $E > V_r$) requires the assumption that $\mu > 1$ in (\ref{Assumption VR}). The derivatives $f_r^{(n)}$ of $f_r$ with respect to $\kappa$ also satisfy an integral equation of the form (\ref{(A)}) obtained by formally differentiating the integral equation for $f_r$ and calculating at each step the inhomogeneous term  $\Phi(\kappa,\sigma;x)\equiv  \Phi(\kappa,x)$ (the parameter $\sigma$ plays no role here). By considering successively $f_r^{(1)}$, $f_r^{(2)}$, $\dots$ and using at each step the bounds of the form (\ref{(D)}) obtained for the derivatives of the lower orders, one finds that the inhomogeneous term in the integral equation for $f_r^{(n)}$ satisfies  (\ref{(B)}) with $\eta(\kappa,\sigma)=c(\kappa)$ and $h(x)=1+x^n$ ($x \geq 0$). The number $c(\kappa)$ is finite if $\kappa > 0$ and $\int_{0}^{\infty} y^n |V(y)-V_r| \, dy < \infty$. So the integral equation for $f_r^{(n)}$ has a unique solution provided that $\mu > n+1$ in (\ref{Assumption VR}). By using the bounds for the functions $f_r^{(j)}$ ($0 \leq j \leq n$), one can also verify a posteriori, proceeding again recursively, that these functions are indeed the derivatives of the Jost solution (using a theorem permitting the interchange of the integral with the derivatives with respect to $\kappa$ in the occuring Volterra equations, \eg Lemma 2 in Chapter XIV of \cite{LA}).

Next let $\kappa,\sigma > 0$ with $|\kappa - \sigma| < 1$ and let $\Psi_{\kappa,\sigma}$ be the difference of the Jost solutions $f_r$ at energy $E_{\kappa}$ and  $E_{\sigma}$: $\Psi_{\kappa,\sigma} = f_r(E_\kappa) -  f_r(E_\sigma)$, where for example $E_{\kappa} = \kappa^2 + V_r$. The function $\Psi_{\kappa,\sigma}$ satisfies an integral equation of the type (\ref{(A)}). By using the inequality $|e^{i\kappa x} - e^{i\sigma x}| \leq 2 |\kappa - \sigma|^{\nu} |x|^{\nu}$ valid for $0 \leq \nu \leq 1$, the addition theorem for the sine function and the bound  (\ref{(D)}) for $f_r(E_{\sigma})$, one obtains an estimate of the form (\ref{(B)}) for the inhomogeneous term $\Phi(\kappa,\sigma;x)$, with $\eta(\kappa,\sigma) = \rho(\kappa,\sigma) |\kappa - \sigma|^{\nu}$ and $h(x)= 1 + |x|^\nu$, where $\rho(\kappa,\sigma)$ is a smooth function (the condition (\ref{(C)}) is satisfied if $\nu < \mu -1$). In view of (\ref{(D)}) we conclude that $f_r(E)$ is H\"{o}lder continuous as a function of $E > V_r$ with any exponent $\nu$ satisfying $\nu < \min\{1,\mu-1\}$. By proceeding again recursively, one can use the integral equation for $f_r^{(n)}(E_{\kappa}) - f_r^{(n)}(E_{\sigma})$ to show that $f_r^{(n)}(E)$ is  H\"{o}lder continuous with any exponent $\gamma < \min\{1,\mu-n-1\}$ if $\mu > n+1$.

\bibliographystyle{apsrev}
\bibliography{Paper-Definitive}

\begin{thebibliography}{52}
\bibitem[{CCC()}]{CCC}
\bibinfo{note}{The characterization of the continuous spectrum follows from
  Corollary II.5 in \cite{Carmona}. The finiteness of the number of bound
  states if $\mu > 2$ is well known, see \eg \cite{Klaus} (combined with the
  min-max principle), \cite{Aktosun} or Theorem 7.5 in \cite{Simon-Trace}.}
\bibitem[{Com({\natexlab{b}})}]{Comment-2}
\bibinfo{note}{We refer to Theorem 8.1(a) of \cite{PSS} for a simple proof.}
\bibitem[{Q()}]{Q}
\bibinfo{note}{A sufficient condition for the finiteness of $N$ is $\mu > 2$.
  The stated conclusion can also be obtained without assuming that $N <
  \infty$, see \eg the reference given in \cite{Comment-3}.}
\bibitem[{XX2()}]{XX2}
\bibinfo{note}{The proof of (\ref{AS1}) in \cite{AC} is given for configuration
  spaces of dimension $n \geq 2$; except for a very minor modification it also
  applies in dimension $n=1$. The $R$-independent term in this asymptotic
  expansion (denoted $A_0/2$ in \cite{AC}) acts as $-id/dE$ in the usual
  representation of $L^2(\mathbb{R})$ that diagonalizes the free Hamiltonian
  $H_0$, and it is easy to check that its action in $\mathcal{H}^{in}$ and in
  $\mathcal{H}^{out}$ is also given by $-id/dE$.}
\bibitem[{NN()}]{NN}
\bibinfo{note}{The terms in $T^{out}_R(\varphi)$ that contain a scalar product
  involving $S_{\ell \ell}\varphi$ and $S_{r \ell}\varphi$ are seen to be of
  the order ${\scriptstyle \cal O}(1)$ as $R\rightarrow \infty$ by using the
  argument given in relation with (\ref{Bound Intersection}).}
\bibitem[{XXX()}]{XXX}
\bibinfo{note}{This can also be seen in the time-dependent formalism: by using
  (\ref{Alternative Exp}) with the explicit expressions for $\psi(x,t)$ given
  in \cite{Cohen}, where $\psi = \Omega^{-}_{\ell}\varphi$, one finds that
  $\tau_\ell(\varphi)=\tau_r(\varphi)=0$ if $\varphi\in\mathcal{H}_+$ has
  energy support (with respect to $H_\ell$) above $V_r$. For wave functions
  with energy support in $(V_\ell,V_r)$ both $\tau_\ell(\varphi)$ and
  $\tau_r(\varphi)$ give a positive contribution to the global time delay.}
\bibitem[{TT()}]{TT}
\bibinfo{note}{An improved version of Mourre theory (Theorems 7.8 and 8.1(a) in
  \cite{PSS}) shows that (\ref{Propagation-Estimate-Strong-Consequence})
  remains valid for functions $f$ satisfying $|f(x)| \leq C(1+|x|)^{-\sigma}$
  for some $\sigma > 1/2$. The existence of the limit in (\ref{Separating LR3})
  for $1 < \mu < 2$ is then obtained by replacing $1+|Q|$ in (\ref{Bound 1+Q}),
  hence also in (\ref{Finite2}), by $(1+|Q|)^{\mu/2}$.}
\bibitem[{Com({\natexlab{c}})}]{Comment-3}
\bibinfo{note}{This requires a rather subtle argumentation which is given in a
  general situation for example in Section 17.2 of the book \cite{BW} (see in
  particular the proof of Theorem 26 and Proposition 27).}
\end{thebibliography}

\end{document}